\documentclass[twocolumn]{aa}
\input psfig.tex

\def\hmpc{\rm \,h^{-1}\,Mpc}

\def\rvsao {{\tt RVSAO}\/}
\def\begc{\begin{center} }
\def\endc{\end{center} } 
\def\begf{\begin{figure} }
\def\endf{\end{figure} }

\usepackage{graphicx}
\begin{document}                                                                



\title{The REFLEX galaxy cluster survey. VIII. Spectroscopic observations
  and optical atlas\thanks{Based on data collected at the
    European Southern Observatory, La Silla, Chile -- The full table
    is available in electronic form at the CDS via anonymous ftp to
    cdsarc.u-strasbg.fr (130.79.128.5) or via
    http://cdsweb.u-strasbg.fr/cgi-bin/qcat?J/A+A/, or
    htpp://www.brera.inaf.it/REFLEX.}} 

   \titlerunning{The REFLEX redshift catalog}

   \author{L. Guzzo$^{1,2}$, 
P.\,\,Schuecker$^{2}$, H.\,\,B\"ohringer$^{2}$, C.A.\,\,Collins$^{3}$,
A.\,Ortiz-Gil$^{4,1}$, S.\,De Grandi\inst{1}, A.C.\,Edge\inst{5},
D.M.\,Neumann\inst{6}, S.\,Schindler\inst{7},
C.\,Altucci$^{1}$, P.A.\,Shaver\inst{8}
} 
                                      
   \authorrunning{Guzzo et al.}

   \offprints{Luigi Guzzo\\ luigi.guzzo@brera.inaf.it}

   \institute{             
    $^{1}$ INAF - Osservatorio Astronomico di Brera, via Bianchi 46, I-23807
 Merate (LC), Italy\\
    $^{2}$ Max-Planck-Institut f\"ur extraterrestrische Physik,
             Giessenbachstra{\ss}e 1, D-85740 Garching, Germany\\ 
    $^{3}$ Astrophysics Research Institute, Liverpool John Moores
 University, Twelve Quays House, Egerton Wharf, Birkenhead CH41 1LD,
 Great Britain\\
    $^{4}$ Observatori Astronomic - Universitat de Valencia,
 Edificio de Institutos de Investigacion, Aptdo. Correos 22085,
    E-46071 Valencia, Spain\\
 $^5$ Physics Department, University of Durham, South Road, Durham DH1 3LE, U.K.\\
 $^6$ CEA Saclay, Service d`Astrophysique, Gif-sur-Yvette, France\\
 $^{7}$ Institute for Astro- and Particle Physics, Universit\"at Innsbruck, 6020 Innsbruck, Austria\\
$^{8}$ European Southern Observatory, D 85748 Garching, Germany\\
}

   \date{ }                         	
   
   \markboth{}{}

   \abstract{ 
     We present the final data from the spectroscopic survey of the
     ROSAT-ESO Flux-Limited X-ray (REFLEX) catalog of galaxy
     clusters. The REFLEX survey covers 4.24 steradians (34\% of
     the entire sky) below a declination of $\delta= +2.5^{0}$ and at
     high Galactic latitude ($|b|>20^\circ$).  The REFLEX catalog
     includes 447 entries with a 
     median redshift of 0.08 and is better than $90\%$ complete to a
     limiting flux $f_X=3\times10^{-12}\rm{erg s}^{-1} \rm{cm}^{-2}$
     (0.1 to 2.4 keV), representing the largest statistically 
     homogeneous sample of clusters drawn from the ROSAT All-Sky Survey
     (RASS) to date.    Here we describe the details
     of the spectroscopic observations carried out at the ESO 1.5~m, 2.2~m,
     and 3.6~m telescopes, as well as the data reduction and
     redshift measurement techniques.  The spectra typically cover the
     wavelength range $3600-7500 \AA$ at a two-pixel resolution
     of $\sim 14 \AA$, and the measured redshifts have a total {\it rms}
     error of $\sim 100$ km  
     s$^{-1}$. In total we present 1406 new galaxy redshifts in 192
     clusters, most of which previously did not have any redshift
     measured.  Finally, the luminosity/redshift 
     distributions of the cluster sample and a comparison to the no-evolution
     expectations from the cluster X-ray luminosity function are 
     presented.   
}

\maketitle

\section{Introduction}\label{INTRO}

Clusters of galaxies represent the largest collapsed objects in the
hierarchy of cosmic structures, stemming from the growth of
fluctuations lying on the high-density tail of the matter density
field (Kaiser 1986).  As such, their number density and evolution are
strongly dependent on the normalization of the power spectrum and the
value of the density parameter $\Omega_M$ (e.g. Borgani \& Guzzo 2001;
Rosati et al. 2002).  In addition, the physics involved in
``illuminating'' clusters and making them visible is in principle 
easier to understand than the various complex processes
connected to the formation and evolution of stars in galaxies (although a
drawback can be that their typical dynamical time is long, comparable
to the Hubble time).  In particular in the X-ray band, where clusters
can be defined and recognised as single objects (not just as a mere
collection of galaxies), observable quantities like X-ray luminosity
$L_X$ and temperature $T_X$ show scaling relations with the total
mass (and thus to the mass of the dark-matter halo, e.g. Evrard et
al. 1996; Allen et al. 2001; Reiprich \& B\"ohringer 2002; Ettori et
al. 2004).  A full comprehension of these scaling relations requires
more ingredients than the simple conversion of gravitational potential
energy into heat during the growth of fluctuations (Kaiser 1986;
Helsdon \& Ponman 2000; Finoguenov et al. 2001; Borgani et al. 2004).
Nevertheless, these relations allow us to use clusters to test the
mass function and the mass power spectrum, respectively via the  
observed cluster X-ray luminosity function (XLF) and clustering,
(e.g. B\"ohringer et al 2002; Pierpaoli et al. 2003;
Schuecker et al. 2003a).

\begin{figure*}          
\begc                                                        
\psfig{figure=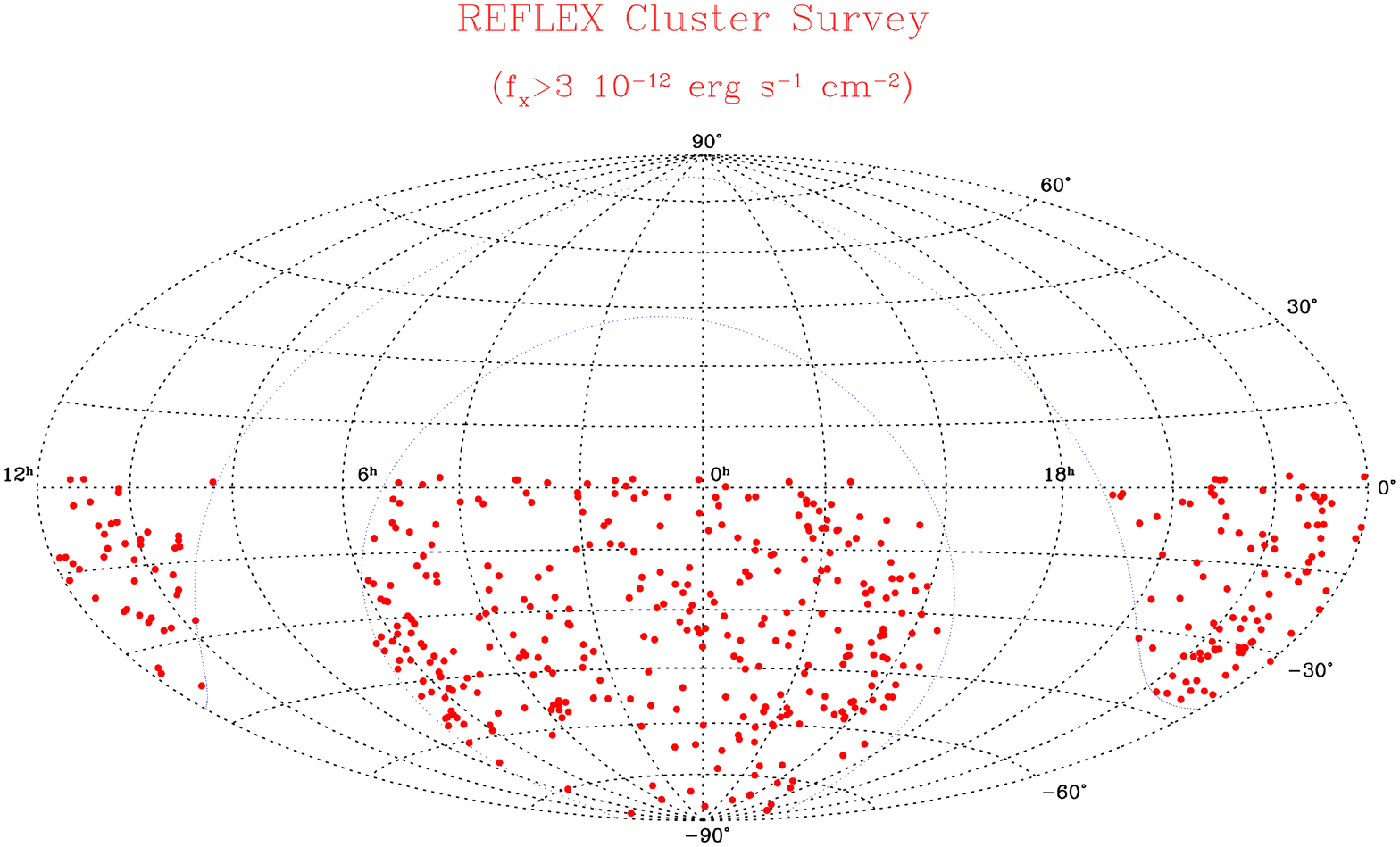,width=18cm}
\caption{The distribution on the sky of the 447 clusters composing the
  REFLEX sample with $f_x> 3 \times 10^{-12}$ erg s$^{-1}$ cm$^{-2}$.
  Note that the apparently blank area around $\alpha\sim1^h$,
  $\sim5^h$, $\delta\sim -70^\circ$ has been excised from the survey,
  corresponding to the Magellanic Clouds.}
\label{aitoff}
\endc
\end{figure*}  

In addition to providing a fairly direct
connection of observed quantities to model (mass-specific)
predictions, X-ray based cluster surveys have further crucial
advantages over optically-selected 
catalogs: first, X-ray emission is proportional to the gas density
squared, and thus is more concentrated and less sensitive to
projection effects than the simple galaxy density profile. Secondly, the
selection function of an X-ray cluster survey is essentially that of a
flux-limited sample, and thus fairly easy to reconstruct.  This is a
crucial feature when the goal is to use these samples for cosmological
measurements that necessarily involve a precise knowledge of the
sampled volume, as it is the case when computing first or second
moments of the density field.

The advent of the ROSAT All-Sky Survey (RASS, Voges et al. 1999) at
the beginning of the 1990's, opened up for the first time the
possibility to construct X-ray cluster samples over wide areas of the
sky.  Optical identification of these clusters was eased by the good
match between the flux limit of the RASS ($\sim 10^{-12} $ erg
s$^{-1}$ cm$^{-2}$ for extended sources) and the depth of the only
wide-area optical imaging available at the time, i.e. the Palomar and
in particular the southern UK-Schmidt sky surveys.
These early X-ray samples included surveys like Hydra (Pierre et
al. 1994), SGP (Romer et al. 1994, Cruddace et al. 2002, 2003), XBACS
(Ebeling et al. 1996), BCS (Ebeling et al. 1998, 2000), RASS-BS (De
Grandi et al. 1999), 
NORAS (B\"ohringer et al. 2000), NEP (Henry et al. 2001, Gioia et
al. 2003).  Some of these early
studies concentrated on X-ray detections of optically-selected
clusters, i.e. typically systems previously identified optically by
Abell (1958) and Abell, Corwin, and Olowin (1989), as notably the
XBACS catalog or the surveys of Burns and collaborators (Burns et
al. 1996; Ledlow et al. 1999). However, some others, as the SGP, BCS and
RASS-BS surveys, were initial steps towards the goal of constructing a
complete, X-ray selected statistical sample covering the whole sky, or
at least the Southern hemisphere where deeper panoramic imaging was
provided by the digitization of the ESO-SRC III-aJ ($b_J$) plates
(through e.g. the Edinburgh-based COSMOS catalog, McGillivray \&
Stobie 1984, or the APM survey, Maddox et al. 1990).

This goal has been achieved with the completion of the REFLEX
(ROSAT-ESO Flux Limited X-ray) cluster survey, whose optical
identification and spectroscopic survey are described here.  REFLEX
combines the X-ray data from the RASS and ESO follow-up optical
observations to construct a complete flux-limited sample of 447
clusters with flux limit $f_x \ge 3 \times 10^{-12}$ erg s$^{-1}$
cm$^{-2}$ (in the ROSAT energy band, 0.1-2.4 keV). It covers the
Southern sky up to $\delta = +2.5^o $, excluding the band of the Milky
Way ($|b_{II}| \le 20^o $) to avoid high NH column densities and
crowding by stars.  For the same reason, the regions of the Magellanic
clouds are also excised from the survey (see Table 1 in B\"ohringer et
al. 2001, Paper I hereafter), totaling an overall area of 13924
$\deg^2$ or 4.24 sr.  The overall sky distribution of REFLEX clusters
is shown in Fig.~\ref{aitoff}.


REFLEX provides the largest statistically complete
X-ray-selected cluster sample to date. The volume of Universe it
probes is bigger than that covered  by any present  galaxy redshift
survey except for the Sloan Digital Sky Survey, which goes to
slightly larger depth but covers about
half the sky area of REFLEX.   We note that the RASS still remains
today the only all-sky X-ray survey  performed with an
imaging X-ray telescope.  We also note that the potential of
the RASS for cluster research has not been fully exploited yet. There
are two ongoing efforts in this direction.  The REFLEX-2 survey is
extending REFLEX to a fainter flux limit of
$f_x=1.8 \times 10^{-12}$ erg s$^{-1}$ cm$^{-2}$.  This sample
will contain more than 400 new clusters, part of which have been
already observed 
spectroscopically during the ESO Key Programme described in this
paper.  Complementarily, the MACS survey (Ebeling et al. 2001; 2007)
aims specifically at identifying all luminous X-ray clusters at
$z>0.3$ still hiding in the RASS, probing an even bigger volume of the
Universe. 

The overall goal of REFLEX has been to map a large volume of the
Universe using clusters, such that the survey could be used both to
measure large-scale structure and as a controlled source for studying
the physical properties of clusters.  These requirements imposed a
high standard to the whole X-ray source selection and identification
process, which is described in detail in B\"ohringer et al. (2004,
Paper II hereafter).  In this paper, we present the data from the
spectroscopic survey conducted with ESO telescopes to identify and
measure the redshifts of REFLEX clusters. In particular, we report all
relevant information on individual galaxy redshift data.  We also
provide (in electronic form), finding charts and optical/X-ray
overlays of the clusters. These allow a first qualitative inspection
of their main morphological properties (as e.g. their
concentration or the presence of a dominant cD galaxy), which we hope
will stimulate further quantitative work on this sample.

The complete REFLEX survey has been used over the last few years to
measure fundamental cosmological quantities in the ``local'' Universe.
These include, among others:

$\bullet$ The cluster X-ray luminosity function (B\"ohringer  et
al. 2002), and from this the mean abundance of clusters.

$\bullet$ The two-point correlation function of the cluster
distribution (Collins et al. 2000).

$\bullet$ The power spectrum of the cluster distribution (Schuecker et
al. 2001, 2002).

$\bullet$ The values of the cosmic mean density of matter $\Omega_M$ and the
power spectrum normalization $\sigma_8$, via the combination of the
above observables (Schuecker et al. 2003a).

$\bullet$ The Gaussianity of the cluster distribution, as described by
Minkowski functionals (Kersher et al. 2001).

$\bullet$ The value of the equation of state parameter of dark energy
$w$ (Schuecker et al. 2003b).

$\bullet$ The relation between cluster velocity dispersions
(measurable for a sub-sample of 170 clusters) and X-ray luminosity
(Ortiz-Gil et al. 2004).

$\bullet$ The cluster-galaxy correlation function (Sanchez et al. 2005).

$\bullet$ The influence of scaling relation uncertainties
on the estimate of cosmological parameters (Stanek et al. 2006).

One further general aspect is that through these measurements REFLEX
provides the currently most robust local ($\left<z\right>\sim 0.05$)
reference frame to which surveys of distant clusters can be safely
compared in search of evolution (e.g. Borgani et al. 2001, Henry 2003,
Stanford et al. 2006).  Finally, the REFLEX catalog has provided the
basis for statistically complete studies of the thermodynamical
properties of the intra-cluster medium and the corresponding scaling
relations. This is the case of the ``REXCESS'' XMM large survey
recently completed (Boehringer et al. 2007). 

The paper is organized as follows: in \S~2 we provide a quick overview
of the selection and identification strategy of the REFLEX survey; in
\S~3 we present the spectroscopic observations and discuss the
observations, data reduction and redshift measurement technique;  in
\S~4 we present the spectroscopic catalog and the related finding
charts and optical overlays; in \S~5 we discuss some properties of the
redshift and luminosity distributions of REFLEX cluster; finally, in
\S 6 we conclude and summarize the content of the paper. 
We adopt a ``concordance'' cosmological
model, with $H_o=70$ km s$^{-1}$ Mpc$^{-1}$, $\Omega_M=0.3$,
$\Omega_\Lambda = 0.7$, and -- unless specified -- quote all X-ray
fluxes and luminosities in the ROSAT [0.1-2.4] keV band.

\section{REFLEX IDENTIFICATION STRATEGY: OVERVIEW}

We summarize here, for completeness, the main stages that led to the
construction of the cluster candidate sample for REFLEX.  A more
comprehensive description can be found in Papers I and II.   

The X-ray data for all sources detected in the RASS at declinations
smaller than $2.5^\circ$ were analysed using the ``Growth Curve
Analysis'' (GCA) method (B\"ohringer et al. 2000), thus re-measuring
their flux and geometrical properties.  The results are used to
produce a flux-limited sample of RASS sources with $f_x \ge 3 \cdot
10^{-12}$ erg s$^{-1}$ cm$^{-2}$.  This redetermination of the fluxes
has been shown to be crucial for extended RASS sources, as are the
majority of REFLEX clusters (Ebeling et al. 1996, De Grandi et
al. 1997, B\"ohringer et al. 2000).  Cluster candidates were then
found correlating all sources with galaxy density enhancements in the
COSMOS optical data base, obtained from digital scans of the UK
Schmidt survey plates at the Royal Observatory Edinburgh (MacGillivray
\& Stobie 1984, Heydon-Dumbleton et al. 1989), with a density
threshold low enough as to guarantee our desired final completeness of
better than 90\%.  This meant accepting a contamination of $\sim 30\%$
by non-cluster sources spuriously associated with fluctuations in the
galaxy background counts.
The procedure ensures that the selection effects introduced by the
optical identification process are minimized and negligible for our
purpose (see also the statistics given in Paper I).  Further tests
provide support that a figure comfortably larger than $90\%$ also
describes the overall detection completeness of the flux-limited
cluster sample in the survey area. 

The resulting candidate list was then carefully checked against the
available X-ray/optical information and with literature data, to
eliminate obvious contaminants prior to the deeper optical follow-up
observation program at La Silla.  The adopted scheme was very
conservative, again accepting a larger contamination (to be cleaned
afterwards by the follow-up observations) to guarantee the highest possible
completeness in the final sample (see Paper I for details). 

\begin{center}
\begin{table*}
\caption{Complete log of the spectroscopic observations.
}
\begin{tabular}{llllllll}
\hline
\hline
 Date/Nights & & Tel. & Spectrograph &  CCD & Grism/
& Disp. & Detector model\\
&  &  & & & Grating & (\AA\/ mm$^{-1}$) & \\
\hline
1992, 27-31 May & (4) & 3.6~m & EFOSC-1 & \#16 &  B300 & 230 & Tek $512\times 512\; 30 \mu$ px \\
1992, 1-5 Jun & (5) & 1.5~m & B\&C & \#27 & \#21 & $130$ & RCA
$512\times 512\; 30 \mu$ px \\
1992, 21-24 Nov & (3) & 2.2~m & EFOSC-2 & \#19  & \#1 & 442 & Thomson
$1024\times 1024$ $19\mu$ px\\
1992, 26-29 Nov & (3) & 3.6~m  & EFOSC-1  & \#26 & B300  & 230 & Tek
$512\times 512\; 27 \mu$ px\\
1993, 16-20 Apr & (4) &  3.6~m & EFOSC-1  & \#26  & B300  & 230 & \\
1993, 13-19 Sep & (6) & 1.5~m  & B\&C  & \#24  & \#27 & $114$ & Ford
$2048\times 2048$ $15\mu$ px\\
1993, 14-17 Dec & (3) & 3.6~m  & EFOSC-1  & \#26  & B300  & 230 & \\
1994, 10-13 Mar & (3) & 1.5~m  & B\&C  & \#24  & \#23 & $129$ & \\
1994, 13-16 Mar & (3) & 2.2~m & EFOSC-2 & \#19  & \#1 & 442 & \\
1994, 6-11 May & (5) & 2.2~m & EFOSC-2 & \#19  & \#1 &  442 & \\
1994, 12-15 May & (3) & 1.5~m  & B\&C  & \#24  & \#27 & $114$ & \\
1994, 9-12 Sep & (3)  & 3.6~m  & EFOSC-1  & \#26  & B300  & 230 & \\
1994, 6-9 Dec & (3) & 3.6~m  & EFOSC-1  & \#26  & B300  &230  & \\
1994, 31 Dec - 1995, 4 Jan & (4) & 2.2~m & EFOSC-2  & \#19   &  \#6 & 137 & \\
1995, 1-7 May & (6) & 2.2~m & EFOSC-2 & \#19   & \#6 & 137  & \\
1995, 25 Sep - 1 Oct & (6) & 2.2~m & EFOSC-2 & \#19  &\#6  &  137 & \\
1995, 20-23 Dec & (3) & 3.6~m  & EFOSC-1  & \#26  & B300  & 230 & \\
1995, 17-20 Dec & (3) & 1.5~m & B\&C  & \#39  & \#23 & $129$ &
Loral/Lesser $2048 \times 2048\; 15 \mu$ px\\
1996, 7-10 Sep & (3) & 3.6~m  & EFOSC-1  & \#26  & B300  & 230 & \\
1996, 10-13 Sep & (3) & 1.5~m & B\&C  & \#24  & \#23 & $129$ & \\
1997, 5-7 Feb & (2) & 1.5~m & B\&C  & \#39  & \#23 & $129$ & \\
1997, 8-11 Feb & (3) & 2.2~m & EFOSC-2 & \#40 &\#6  & 136 &
Loral/Lesser $2048 \times 2048\; 15 \mu$ px\\
1997, 1-2 Jun & (1) & 3.6~m  & EFOSC-1  & \#26  & B300  & 230 & \\
1997, 2-6 Jun & (4) & 1.5~m & B\&C  & \#39  & \#23 & $129$ & \\
1997, 29 Sep - 2 Oct & (3) & 3.6~m  & EFOSC-1  & \#26  & B300  & 230 & \\
1998, 30 Jan - 1 Feb & (1) & 3.6~m  & EFOSC-2  & \#40  & \#11 & 136 & \\
1998, 17-20 Sep & (3) & 1.5~m & B\&C & \#39  & \#23 & $129$ & \\
1998, 20-22 Sep & (2) & 3.6~m  & EFOSC-2  & \#40  & \#11 & 136 & \\
1999, 17-20 May & (3) & 3.6~m  & EFOSC-2  & \#40  & \#11 & 136 & \\
\hline
\hline
\label{log}
\end{tabular}
\end{table*}
\end{center}
The follow-up optical observations of REFLEX clusters were started at
ESO in May 1992.  With an ``ESO Key Programme'' status, the
survey obtained an overall allocation of 90 nights, distributed among
the ESO 3.6~m, 2.2~m and 1.5~m telescopes.  A few additional nights
were further obtained at the end of the project to partly compensate
for the time lost due to bad weather.  The complete observing log of
the survey is presented in Table~\ref{log}.

The goal of these observations was twofold: a) obtain a definitive
identification of ambiguous candidates; b) obtain a measurement of the
mean cluster redshift.  First, a number of candidate clusters required
direct CCD imaging and/or spectroscopy to be safely included in the
sample.  For example, candidates characterised by a poor appearance on
the Sky Survey IIIa-J plates, with no dominant central galaxy or
featuring a seemingly point-like X-ray emission had to pass further
investigation.  In this case, either the object at the X-ray peak was
studied spectroscopically, or a short CCD image plus a spectrum of the
2-3 objects nearest to the X-ray peak were taken. This operation was
preferentially scheduled at the two smaller telescopes (1.5~m and
2.2~m, see below).
In this way, a few AGN's were discovered.  When the overall
information available (e.g. X-ray hardness ratio, source shape) was
consistent with the AGN dominating the emission, the
corresponding candidate was rejected from the main list. This is
described in full detail in Paper II, where also a list of the
 more uncertain or ambiguous cases in the REFLEX catalog is
 presented and discussed thoroughly.

For the {\it bonafide} clusters, the final goal of the optical observations
was then to secure a reliable redshift.  The observing strategy was
designed as a compromise between the desire of having several
redshifts per cluster, coping with the multiplexing limits of the
available instrumentation, and the large number of clusters to be
measured.  Previous experience on the similar Edinburgh/Milano Survey
of EDCC clusters (Collins et al.  1995), had shown the importance of
not relying on just one or two galaxies to measure the cluster
redshift, especially for clusters without a dominant cD
galaxy.  However, the additional information provided by the detection
and localisation of X-rays makes the issues of projection -- 
that make multiple member redshifts vital for optical samples -- much
less severe here\footnote{In fact, the data from the REFLEX survey itself
  show exactly this: X-ray emission provides an extremely
  good guidance 
  towards targeting galaxies which have a high probability to be
  cluster members (see also Crawford et al. 1999).  This allows the
  cluster mean redshift to be 
  constrained with fewer objects than for a ``blind'' survey of
  optically-selected clusters, as the EDCC. }.

EFOSC1 in MOS mode was a perfect instrument for getting quick redshift
measurements for 10-15 galaxies at once, but only for systems that
could reasonably fit within the small field of view of the instrument
(5.2 arcmin side in imaging with the Tektronics CCD \#26, but less than 3
arcmin for spectroscopy in MOS mode, due to hardware/software
limitations in the making of the MOS masks).  This feature 
made this combination useful only for clusters above z$\sim0.1$,
i.e. where at least the core region could be accommodated within the
available area (a core radius of $0.1 \hmpc$ is seen under an angle of
1.3 arcmin at such redshift, in the adopted cosmology).

The other important aspect of this instrumental set-up is that in
several cases, after removal of background/foreground objects one is
still left with 8-10 galaxy redshifts within the cluster, by which a
first estimate of the cluster velocity dispersion can be attempted.
This has been done, complementing the data described here with
literature redshifts, for a sub-sample of 170 REFLEX clusters,
allowing us to study the scaling relation between cluster velocity
dispersion and X-ray luminosity (Ortiz-Gil et al. 2004).

At lower redshifts, doing efficient multi-object spectroscopy work
on cluster fields would have required a MOS spectrograph with a larger
field of view, i.e. 20-30 arcminutes diameter.  One possible choice
could have been the formerly available ESO fibre spectrograph Optopus
(Avila et al. 1989), 
but its efficiency in terms of numbers of targets observable per night
was too low for covering the several hundred clusters we had in our
sample.  We found the best solution was to split the work between the
1.5~m and 2.2~m telescopes.  Clearly, this required accepting some
compromise in our initial goal of having multiple redshifts for each
cluster.  As discussed in Paper II, about half of the cluster
redshifts are measured with 5 or more member galaxies, but 42 of them
featuring only one galaxy redshift.  Most of these cases come from the
literature, and the available telescope budget did not allow for a
re-determination of these values.  For most of these cases, however,
the reliability of these single redshift as estimators of the mean
systemic redshift is high, as they refer to the brightest cluster
galaxy at the centre of X-ray emission.  Indeed, as mentioned above,
the coupling of the galaxy positions with the X-ray contours is of
strong help in indicating which galaxies have the highest probability
to be cluster members.

During 8 years of work, we have observed spectroscopically a total of
about 500 cluster candidates, collecting over 3200 galaxy spectra.
In this paper, we present the spectroscopic data belonging
to the current, published  REFLEX sample with $f_x > 3 \times 10^{-12}$ erg
s$^{-1}$ cm$^{-2}$, which include $\sim 1500$ spectra.  The remaining 
spectra belong to clusters extending to fainter fluxes, which will form
part of a deeper REFLEX-2 sample reaching to     
$f_x=1.8 \times 10^{-12}$ erg s$^{-1}$ cm$^{-2}$. 

\section{SPECTROSCOPY}

\subsection{Observations}

As detailed in Table~\ref{log}, all spectroscopic observations were
performed at the ESO La Silla observatory, using the 3.6~m, 2.2~m and
1.5~m telescopes. In the following, we describe the instrumental
set-ups and main data properties for each of them.

\subsubsection{EFOSC1/2@3.6~m Observations}

\begin{figure*}                                                                  
\resizebox{9truecm}{!}{\includegraphics{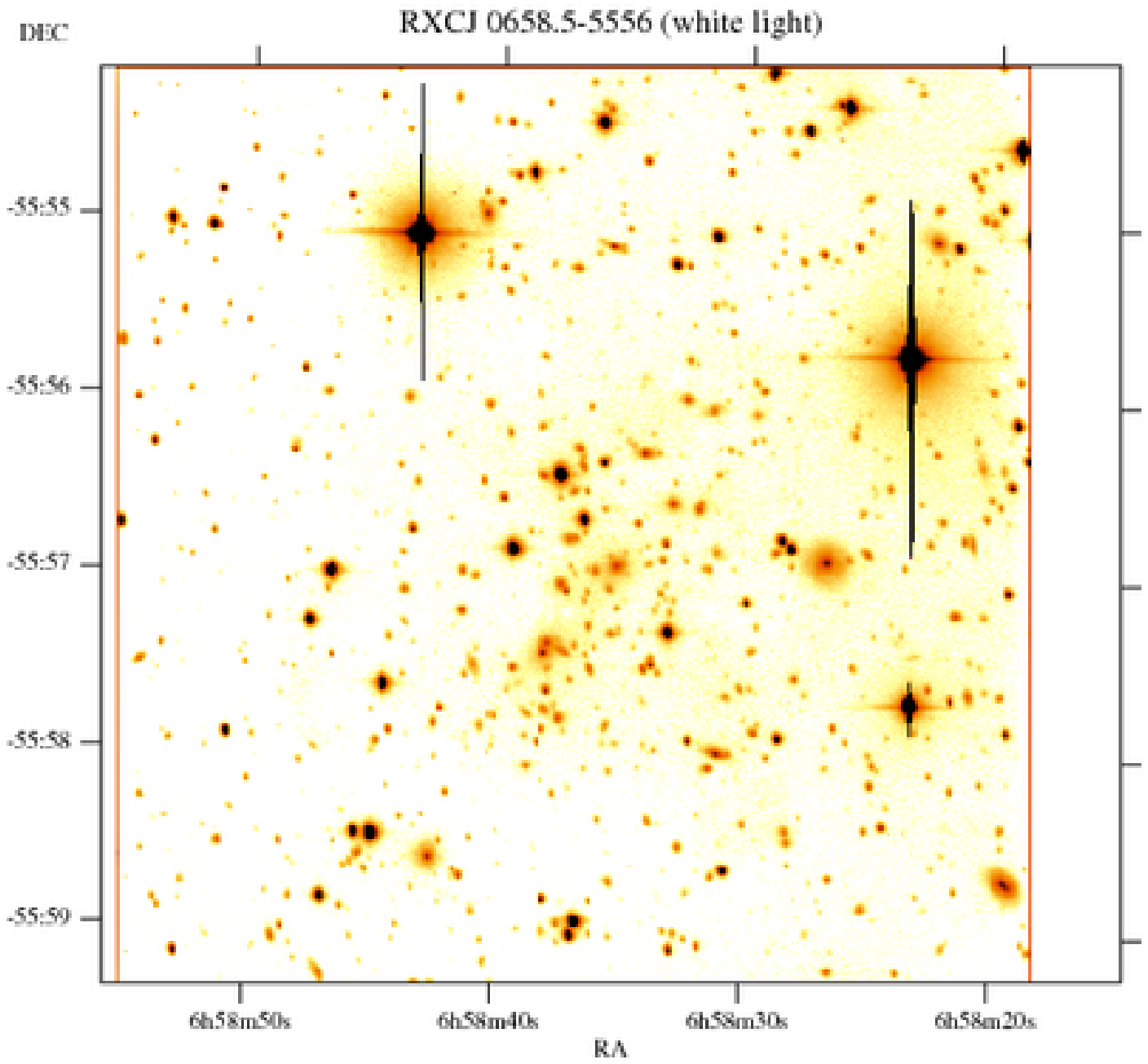}}
\resizebox{9truecm}{!}{\includegraphics{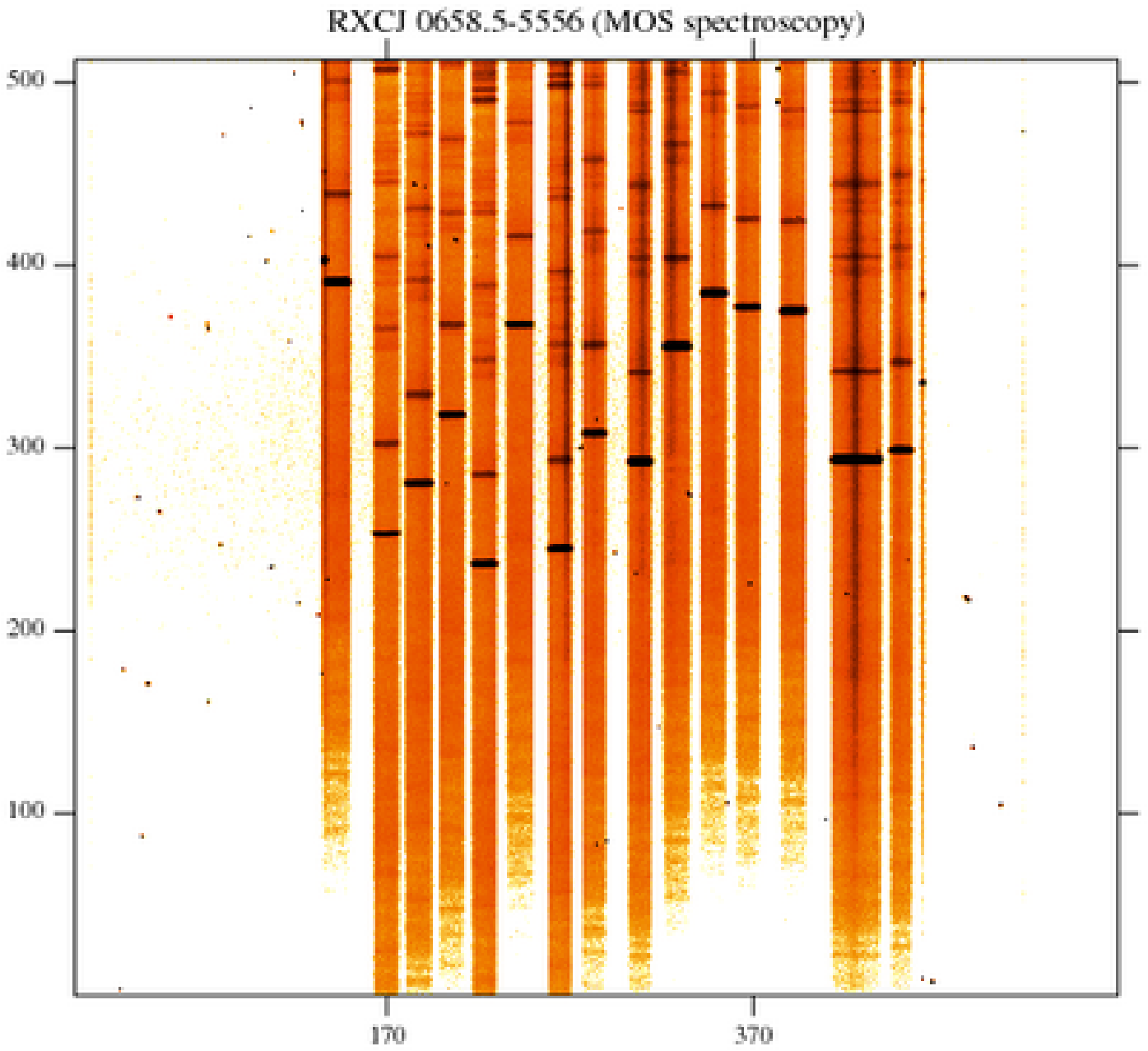}}
\caption{Example of a direct image and spectral lay-out for a cluster
observed in MOS mode with EFOSC-1 at the 3.6~m telescope. {\it Left:}
Direct image in white light of RXCJ0658.5-5556, a luminous REFLEX cluster at
$z=0.2965$ (see also Fig.~\ref{overlays}). {\it Right:}  The resulting
MOS frame,  
showing the set of two-dimensional spectra corresponding to each target
galaxy in the mask.  The dispersion runs along the vertical direction.
Each strip shows the sky spectrum (dark horizontal lines)
together with the fainter galaxy spectrum. }
\label{mos-frame}
\end{figure*}  
About 80\% (in terms of number of spectra) of all the REFLEX survey
observations were carried out using the 3.6 m telescope with the ESO
Faint Object Spectrograph and Camera (EFOSC) in its two incarnations -
EFOSC1 and EFOSC2.  The EFOSC instruments are high-efficiency
transmission spectrographs, with multi-object spectroscopic capability
(MOS) and fast switching to imaging mode. This latter feature allows
very accurate slit positioning on faint objects.  At the time of
completing this paper (2008), EFOSC-2 is still actively used at the 3.6~m
telescope.

Table~\ref{log} shows how during the development of the survey
different detectors were installed on the EFOSCs, following the
evolution of CCD's technology.
The EFOSCs were mostly used in MOS mode, which entailed producing
aluminium masks of the cluster fields, on which slitlets of 5-30
arcsec length were carved following
a direct image taken with the same instrument.  The
masks were then inserted into free positions in the aperture wheel of
the spectrograph.  The width of the slits was always of 2-arcsec, the
same width used for single-slit observations.  Depending on the
available CCD-grism combination, which varied during the survey, we
worked at dispersions ranging between 130 and $230\,\rm{\AA/mm} $,
usually aiming at a wavelength coverage between $3600\,\rm{\AA}$ and
$7500\,\rm{\AA}$.  Most of the 3.6~m observations were performed using
EFOSC-1 with the B300 grism at $230\,\rm{\AA/mm} $ and a Tektronics
$512 \times 512$ chip, yielding a resolution of $6.9\,\rm{\AA}$ per
pixel.  This corresponds to a 
spectral resolution (as measured on a purely instrumentally broadened
line) of $\sim 2$ pixels FWHM, providing radial velocity errors well
below 100 ${\rm km\,s}^{-1}$ for good S/N ratio spectra obtained from
two consecutive exposures of $10-15$ minutes each (see \S~\ref{errors}
for details). On average each mask contained $15-20$ slitlets, over
the available $5.2\times5.2$ arcmin$^2$ field of view.  Standard
calibration observations were collected, as
discussed in more detail in \S~\ref{data-red}.
A number of single-slit observations were also carried out at the 3.6~m
telescope, with a similar set-up, especially near the end of the
survey, targeting some of the most distant clusters in the sample.

 
\subsubsection{EFOSC2@2.2~m and B\&C@1.5~m Observations}

A fraction of the spectroscopic observations were carried out in
single-slit mode using either EFOSC-2 at the $2.2\,\rm{m}$ telescope,
(before this instrument was moved to the 3.6~m telescope in January
1998), or the 1.5 m telescope, equipped with a classical Boller \&
Chivens spectrograph coupled with an RCA (1992) or Ford (1993 onwards)
CCD.  The latter had very poor response in the blue range, i.e. below
4500 \AA, where some of the most interesting absorption lines
(e.g. Calcium H and K, G-band) fall for galaxies at low
redshift. Thus, the 1.5~m telescope was essentially reserved to 
observe 
the brightest members (up to $m_B\sim 17.5$) in the more nearby
clusters of the sample, and played a minor role in the overall
redshift survey.  The spectral setup was similar to that adopted for
the EFOSC spectrographs (grating \#21, 172 \AA\/ mm$^{-1}$, blaze
angle $\theta=6^\circ\;54^\prime$).

\subsection{Data Reduction}
\label{data-red}

\begin{figure*}                                                                  
\resizebox{9truecm}{!}{\includegraphics{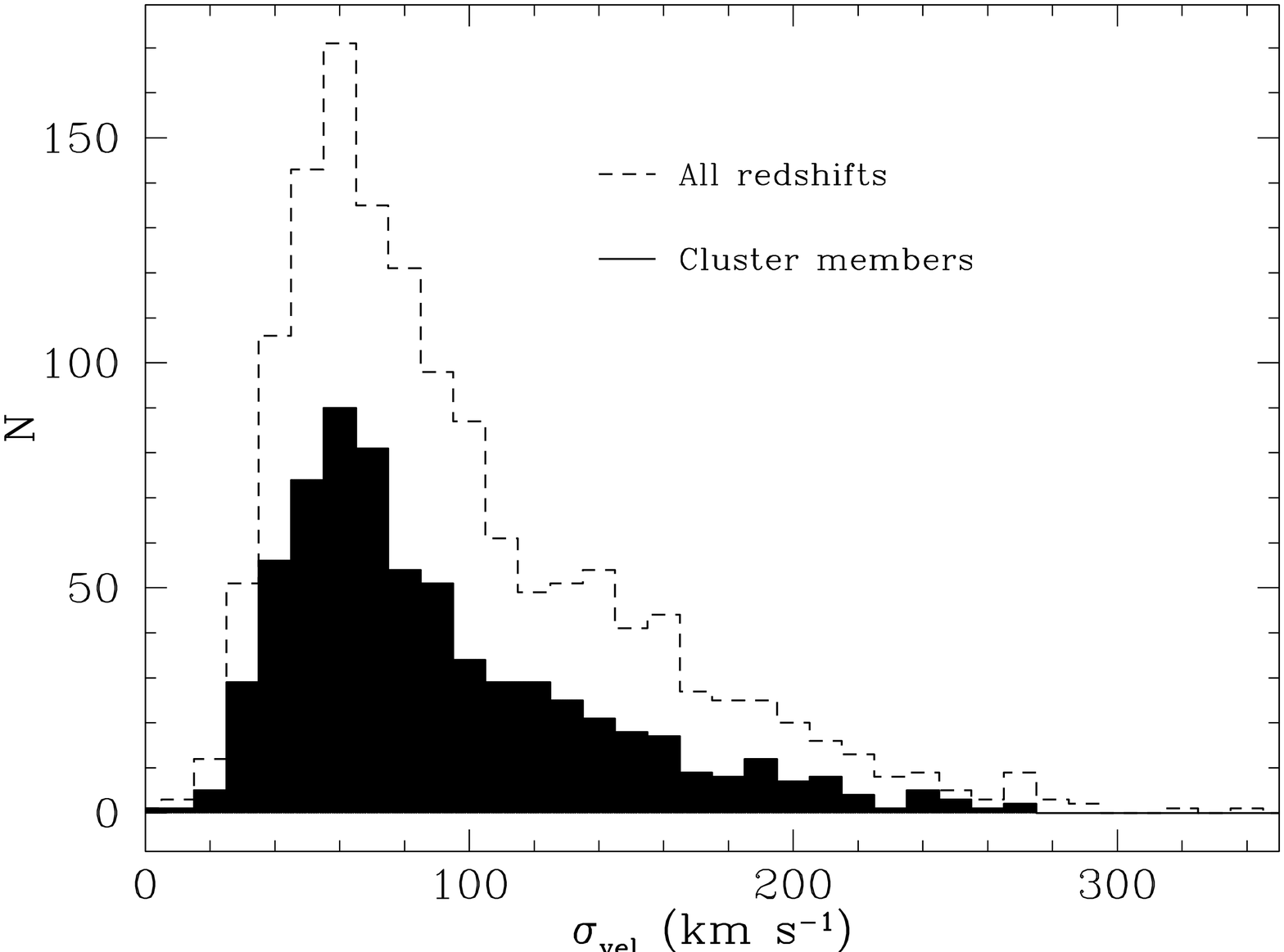}}
\resizebox{9truecm}{!}{\includegraphics{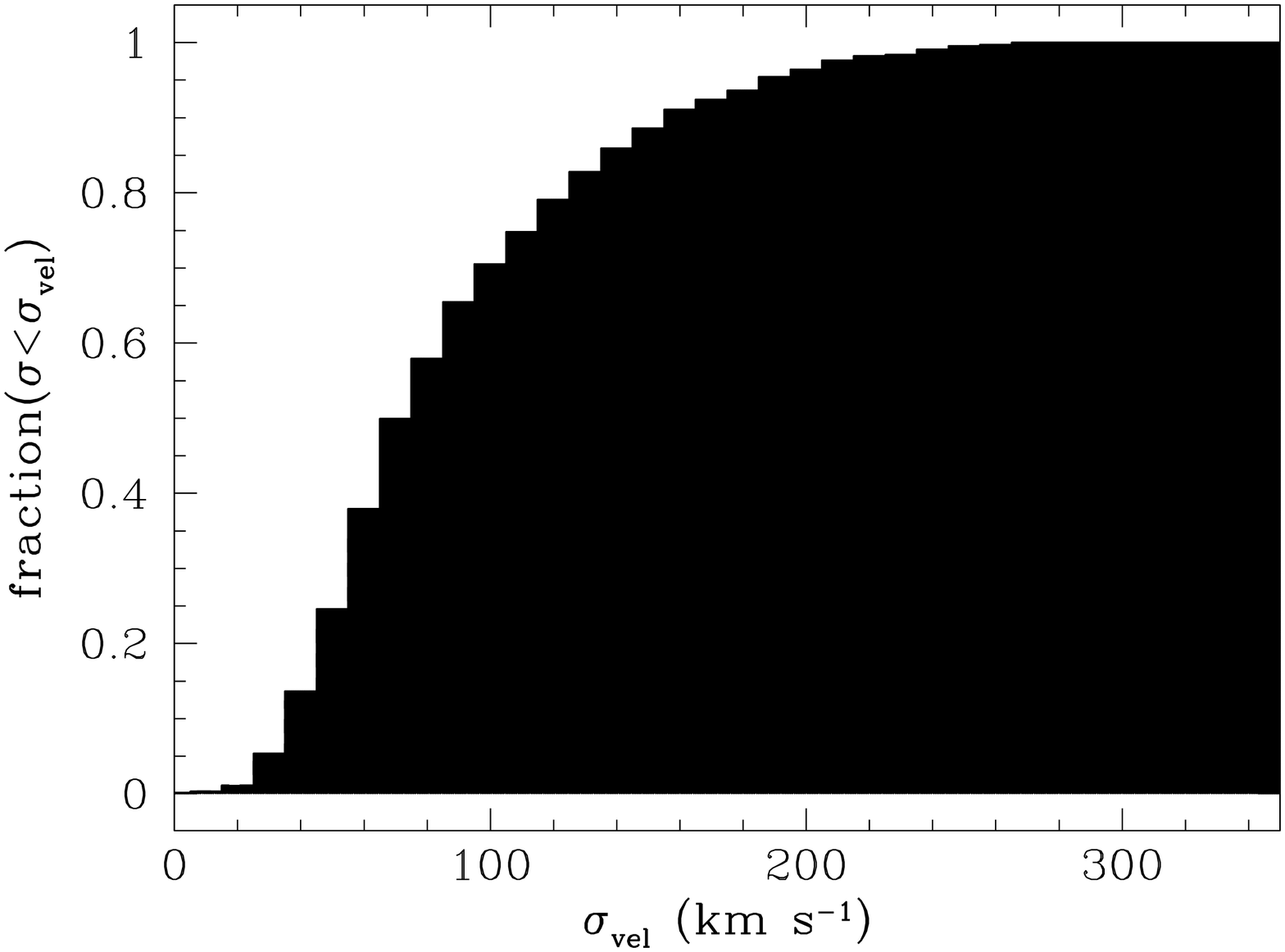}}
\caption{{\it Left:} Differential distribution of the errors on galaxy
  radial velocities in the REFLEX survey, as estimated by \rvsao.  The
  filled histogram corresponds to considering only cluster member
  galaxies (i.e. those that were actually used to compute mean cluster
  redshifts). Collins et al. (1995)
    showed that total redshift errors amount to typically between 1
    and 2 times the internal error estimate by \rvsao\/ , depending
    inversely on
    the SNR of the spectrum. 
  Accounting for this, and considering that all spectra around or
    to the left of the peak of the distribution have very good SNR, we
    conclude that the typical total error on single galaxy
    redshifts remains below 100 km s$^{-1}$.  This assures
    that the dominant source of uncertainty in the mean cluster
  redshift will be the intrinsic 
  velocity dispersion of the cluster and not a big uncertainty
    on single galaxy measurements. 
{\it Right:} Corresponding cumulative error distribution.
}
\label{err_hist}
\end{figure*}  
The data were reduced using the MIDAS (for data prior to 1995) and
IRAF spectroscopic packages, using either custom-built programs (for
MIDAS) or -- for the bulk of the data -- the IRAF specific set of
procedures ({\tt TWODSPEC/APEXTRACT}).  The set of operations
performed on the available long-slit or MOS spectroscopic CCD frames
followed the usual standard procedures, and was essentially the same
in both environments.  For two observing runs, we repeated the full
data reduction using both packages and a direct comparison of the
calibrated spectra showed differences well below our typical radial
velocity errors ($<30$ km s$^{-1}$).  In the following, we shall limit
ourselves, for simplicity, to the IRAF version of the reduction
pipeline which in the end was used for most of the spectra, describing
its various phases.
\\
$\bullet$ {\bf CCD frame inspection, quality check and
  standardization}.  These operations included in particular checks
for: (a) Possible systematic time dependences of the average bias. (b)
Possible shifts of the sky lines during the observing run.  Sky line
positions were also checked after wavelength calibration (see below).
(c) Similarly, possible shifts of He/Ar/Ne comparison lines at
different times during the observing run.  All available science and
calibration frames were then trimmed to a common size, to eliminate
spurious extra borders and overscan regions.
\\
$\bullet$ {\bf Bias and flat-field corrections}.  Multiple sets of
bias frames were regularly collected during each observing night and
combined through a $3\sigma$-clipping algorithm ({\tt ZEROCOMBINE}),
to produce a single, two-dimensional bias frame for that night.
In general, the bias frames from the 1.5~m, from the 2.2~m, and from
the 3.6~m telescope showed two-dimensional structures at the $<0.5$
percent level (RMS) which are removed by this procedure.  To
flat-field our spectra only dome flats were typically observed
in day time during each run, given that we were not aiming
at precise spectrophotometry.  Median flats were constructed for each
run using {\tt IMCOMBINE}.
The number of effectively used flat field exposures for each
Single-Slit spectroscopic run ranged between 7 and 40.  
In practice, we observed virtually no effect when flat-fielding data
from the 1.5~m and the 3.6~m observations, while this operation was
crucial for most of the data collected at the 2.2~m.  Only in the very
last 2.2~m+EFOSC2 run (February 1997) a new CCD (\#40) was installed,
eliminating this problem (EFOSC2 was then moved later in that year to
the 3.6~m telescope, where we then performed most of the subsequent
observations).  
We concluded that there was no gain in flat-fielding the MOS
observations collected at the 3.6~m telescope with EFOSC1 (CCD \#26)
and EFOSC2 (CCD \#40).  
Finally the two (or more) science exposures available for every spectroscopic 
observation were averaged together with {\tt IMCOMBINE}, after
appropriate scaling and weighting by the the exposure time.  This
removed very effectively most of the cosmic ray events.
\\
$\bullet$ {\bf Science and comparison spectra
  extraction}. Two-dimensional spectra corresponding to each slit were
then extracted following possible curvature of the spectrum. These
were then reduced to 1-D sky-subtracted spectra using proper sky
background regions in the slit.  All operations were performed within
the {\tt APALL/APEXTRACT} environment of IRAF.
Corresponding 1D calibration spectra were also extracted at exactly
the same positions from all the available lamp exposures associated
with the target frame.  These were typically two He-Ar arc frames,
observed before and after the science exposure.  At the 3.6~m
telescope, our direct tests for instrument flexures using 5 strong He
lines at 7 extreme telescope positions showed an $rms$ shift $<0.12$
pixels, corresponding to 0.744 \AA, i.e. 45 km s$^{-1}$.  We concluded
that EFOSC flexures over the time of one typical 3.6~m observation
($\sim 30$ min) were negligible, and subsequently used only one
calibration lamp.  This
was not the case at the 1.5~m and 2.2~m telescopes, as shown by
measurements by the ESO staff. For these data, we used both arcs
observed before and after the science exposures to compute a
time-averaged set of reference lines, as feasible within the IRAF
procedures.
\\
$\bullet$ {\bf Wavelength calibration}. The whole operation was
automatized through the available IRAF procedures ({\tt
  IDENTIFY/REIDENTIFY}).  In general, an accurate pixel-to-wavelength
transformation was determined for the first spectrum of either a full
night of long-slit spectroscopy or a single MOS exposure.  The
residuals were directly inspected and discrepant arc line
identifications eliminated. The procedure was iterated until a
satisfactory {\it rms} was reached. The relation was then applied to
the science spectrum using {\tt DISPCOR}.
Typical {\it rms} wavelength calibration errors 
ranged between $\sim 0.3 {\rm\AA}$ (for the majority of spectra,
e.g. those taken at the 3.6~m telescope), to $\sim 1 {\rm\AA}$ for
lower resolution spectra as those obtained at the 2.2~m telescope with
grism \#1.  
All other spectra in a night-series of long-slit observations were
then calibrated by using the first solution as a guess, using {\tt
  REIDENTIFY}.  For MOS spectra, however, where large shifts in the
zero point between adjacent spectra are normal (see
Fig.~\ref{mos-frame}), the position of a bright Helium line was used
to provide an approximate zero-point shift for each spectrum. This was
done through a custom-developed script, and provided the first-guess
to calibrate all spectra in a MOS frame with the usual procedure.  The
quality and consistency of the final calibration was counter-checked a
posteriori by measuring the position of the three brightest sky lines
([OI] $\lambda 5577$,  NaI $\lambda 5891$, and [OI] $\lambda 6300$),
on the calibrated sky spectrum of each extracted science slit. This
allowed us to spot and correct a few pathological cases. 
\\
$\bullet$ {\bf Final cleaning and heliocentric corrections}. Before
feeding the 1D wavelength-calibrated spectra to the cross-correlation
analysis, a number of quality checks and final corrections were
performed.  These include cleaning of bright sky line residuals (via
both an automatic cleaning routine plus visual inspection),
computation of heliocentric corrections using the {\tt RVCORRECT}
package (typically smaller than 30 km s$^{-1}$). Before
actually feeding the spectra to the cross-correlation routine,
emission lines were removed automatically, as only absorption-line
templates were used for the measurement.  Emission-line redshifts were
estimated separately, using the specific routine {\tt EMSAO}.

\subsection{Redshift Measurements}

\subsubsection{Cross-Correlation Technique}

Galaxy redshifts were estimated from the 1D calibrated spectra, using
the classical cross-correlation technique described in detail by Tonry
\& Davis (1979).  This is implemented within the IRAF environment
through the package \rvsao ~(Kurtz \& Mink 1998).  The basis of the
technique is the cross-correlation of the observed galaxy spectrum
with a model or template spectrum.  This is performed by taking the
Fast Fourier Transform of the two spectra, multiplying them together
and then transforming back the result to get the Cross-Correlation
Function (CCF), whose highest peak is related to the radial velocity
difference between the two spectra.  Before actually starting this
machinery, the two spectra are rebinned into logarithmic bins, so that
the relative redshift becomes a linear shift.  Then, a number of
operations are performed on the spectra, in order to improve the
signal-to-noise of the final cross-correlation function.  These
include continuum subtraction, apodizing and bandpass filtering.  All
these operations are performed inside the {\tt XCSAO} routine of
\rvsao.  We tested several combinations of the command parameters to
find the most appropriate set for our spectra.  For example, the
values for the low- and high-frequency cut-offs of the bandpass filter
are specific for the kind of data being used, and optimal values were
chosen after experimenting, as to maximize the significance of the
CCF.  Filtering is important in order to eliminate both the low
frequency spurious components left by the subtracted continuum, and
the high frequency binning noise. Also, we tested that the redshift
estimate was quite insensitive to the exact binwidth (corresponding to
2048 or 4096 bins) chosen in the rebinning.  The peak of the CCF was
fit by a quadratic polynomial, determining the wavelength shift from
its position, and providing an estimate of the uncertainty from its
width.



\newcommand\cola {\null}
\newcommand\colb {&}
\newcommand\colc {&}
\newcommand\cold {&}
\newcommand\cole {&}
\newcommand\colf {&}
\newcommand\colg {&}
\newcommand\colh {&}
\newcommand\coli {&}
\newcommand\colj {&}
\newcommand\colk {&}
\newcommand\coll {&}
\newcommand\eol{\\}
\newcommand\extline{&&&&&&&&&&\eol}

\begin{table*}
  \centering
  \caption[]{Sample page from the full catalogue of REFLEX galaxy redshifts 
(full table available in the electronic version of the journal and at {\tt
  http://www.brera.inaf.it/REFLEX}).   
} 
 
\begin{tabular}{ccclclllclll}
\hline\hline
\cola REFLEX\colb RA\colc DEC\cold Type\cole Cluster \colf cz$_{abs}$\colg err\colh R\coli cz$_{em}$\colj Date\colk Tel\coll Notes\eol
\cola target \colb  {\tiny(hh:mm:ss)} \colc  {\tiny(dd:mm:ss)} \cold  \cole member \colf {\tiny (km/s)}\colg \colh \coli  {\tiny (km/s)}\colj \colk \eol
\hline\hline
\extline
\cola RXCJ0014.3-6604\colb     0:13:56.88\colc    -66:04:14.5\cold galaxy\cole -
\colf   82756\colg  142\colh  3.0\coli        \colj 01-Nov-92\colk 3.6m\coll low SNR\eol
\cola RXCJ0014.3-6604\colb     0:14:00.24\colc    -66:04:59.9\cold galaxy\cole -
\colf  132089\colg   42\colh  3.3\coli 37260  \colj 01-Nov-92\colk 3.6m\coll low SNR\eol
\cola RXCJ0014.3-6604\colb     0:14:01.27\colc    -66:04:39.4\cold galaxy\cole -
\colf  184316\colg  123\colh  2.9\coli        \colj 01-Nov-92\colk 3.6m\coll low SNR\eol
\cola RXCJ0014.3-6604\colb     0:14:04.94\colc    -66:05:38.0\cold galaxy\cole -
\colf   36817\colg   87\colh  3.2\coli        \colj 01-Nov-92\colk 3.6m\coll low SNR\eol
\cola RXCJ0014.3-6604\colb     0:14:05.93\colc    -66:05:35.5\cold galaxy\cole +
\colf   47946\colg   90\colh  3.6\coli        \colj 01-Nov-92\colk 3.6m\coll low SNR\eol
\cola RXCJ0014.3-6604\colb     0:14:05.35\colc    -66:04:21.0\cold galaxy\cole +
\colf   48051\colg   74\colh  6.9\coli        \colj 01-Nov-92\colk 3.6m\coll ---\eol
\cola RXCJ0014.3-6604\colb     0:14:09.14\colc    -66:04:10.6\cold galaxy\cole -
\colf   21356\colg   76\colh  2.8\coli        \colj 01-Nov-92\colk 3.6m\coll low SNR\eol
\cola RXCJ0014.3-6604\colb     0:14:11.66\colc    -66:04:41.9\cold cD gal.\cole +
\colf   45963\colg   59\colh  8.9\coli        \colj 01-Nov-92\colk 3.6m\coll ---\eol
\cola RXCJ0014.3-6604\colb     0:14:15.53\colc    -66:05:38.8\cold galaxy\cole -
\colf   25197\colg   93\colh  3.0\coli        \colj 01-Nov-92\colk 3.6m\coll low SNR\eol
\cola RXCJ0014.3-6604\colb     0:14:16.20\colc    -66:04:21.7\cold galaxy\cole +
\colf   47729\colg   72\colh  5.9\coli        \colj 01-Nov-92\colk 3.6m\coll ---\eol
\cola RXCJ0014.3-6604\colb     0:14:19.58\colc    -66:04:52.3\cold star\cole -
\colf    -251\colg   43\colh  8.4\coli        \colj 01-Nov-92\colk 3.6m\coll ---\eol
\cola RXCJ0014.3-6604\colb     0:14:22.08\colc    -66:04:57.7\cold galaxy\cole +
\colf   48638\colg   72\colh  3.5\coli        \colj 01-Nov-92\colk 3.6m\coll low SNR\eol
\cola RXCJ0014.3-6604\colb     0:14:26.95\colc    -66:04:17.1\cold galaxy\cole -
\colf   74613\colg   94\colh  3.7\coli        \colj 01-Nov-92\colk 3.6m\coll low SNR\eol
\cola RXCJ0014.3-6604\colb     0:14:28.92\colc    -66:04:44.1\cold star\cole -
\colf      12\colg   77\colh  4.6\coli        \colj 01-Nov-92\colk 3.6m\coll ---\eol
\extline
\cola RXCJ0017.5-3509\colb     0:17:36.05\colc    -35:10:47.3\cold galaxy\cole +
\colf   29455\colg  194\colh  6.9\coli        \colj 16-Sep-93\colk 1.5m\coll ---\eol
\cola RXCJ0017.5-3509\colb     0:17:34.87\colc    -35:11:00.2\cold galaxy\cole +
\colf   28071\colg  100\colh 11.1\coli        \colj 16-Sep-93\colk 1.5m\coll ---\eol
\cola RXCJ0017.5-3509\colb     0:17:31.87\colc    -35:11:54.6\cold galaxy\cole +
\colf   29653\colg  151\colh  9.8\coli        \colj 16-Sep-93\colk 1.5m\coll ---\eol
\extline
\cola RXCJ0027.3-5015\colb     0:27:27.60\colc    -50:14:31.9\cold galaxy\cole -
\colf   67964\colg  127\colh  3.6\coli        \colj 19-Sep-93\colk 1.5m\coll low SNR\eol
\cola RXCJ0027.3-5015\colb     0:27:23.45\colc    -50:14:40.6\cold galaxy\cole -
\colf   38130\colg  220\colh  4.7\coli   37795\colj 19-Sep-93\colk 1.5m\coll low SNR\eol
\cola RXCJ0027.3-5015\colb     0:27:20.66\colc    -50:14:46.3\cold galaxy\cole +
\colf   43399\colg   71\colh 13.4\coli        \colj 19-Sep-93\colk 1.5m\coll ---\eol
\extline
\cola RXCJ0042.1-2832\colb     0:41:56.26\colc    -28:30:43.9\cold galaxy\cole -
\colf   16878\colg  267\colh  3.3\coli        \colj 01-Nov-92\colk 2.2m\coll ---\eol
\cola RXCJ0042.1-2832\colb     0:41:58.99\colc    -28:31:08.0\cold galaxy\cole -
\colf   15892\colg  176\colh  5.3\coli        \colj 01-Nov-92\colk 2.2m\coll ---\eol
\cola RXCJ0042.1-2832\colb     0:42:02.42\colc    -28:31:34.3\cold galaxy\cole -
\colf   16492\colg  232\colh  4.3\coli        \colj 01-Nov-92\colk 2.2m\coll ---\eol
\cola RXCJ0042.1-2832\colb     0:42:08.90\colc    -28:32:08.5\cold galaxy\cole -
\colf   16135\colg  168\colh  5.5\coli        \colj 01-Nov-92\colk 2.2m\coll ---\eol
\cola RXCJ0042.1-2832\colb     0:42:11.78\colc    -28:32:35.5\cold galaxy\cole -
\colf   16253\colg  154\colh  5.5\coli        \colj 01-Nov-92\colk 2.2m\coll ---\eol
\cola RXCJ0042.1-2832\colb     0:42:14.86\colc    -28:32:55.7\cold galaxy\cole -
\colf   15572\colg  275\colh  3.3\coli        \colj 01-Nov-92\colk 2.2m\coll ---\eol
\cola RXCJ0042.1-2832\colb     0:42:08.26\colc    -28:32:07.8\cold galaxy\cole +
\colf   33305\colg   75\colh  9.0\coli        \colj 01-Nov-92\colk 2.2m\coll ---\eol
\cola RXCJ0042.1-2832\colb     0:42:08.90\colc    -28:32:08.5\cold cD gal.\cole +
\colf   32296\colg   86\colh 10.1\coli        \colj 01-Nov-92\colk 2.2m\coll ---\eol
\cola RXCJ0042.1-2832\colb     0:42:10.42\colc    -28:32:08.9\cold galaxy\cole +
\colf   32147\colg  172\colh  3.5\coli        \colj 01-Nov-92\colk 2.2m\coll ---\eol
\extline
\hline
\end{tabular}
\label{tab:gal_z}
\end{table*}



\subsubsection{Template Spectra}

At the core of the cross-correlation technique is the comparison of
the object spectrum with a model spectrum of known radial velocity and
ideally infinite S/N ratio, the {\it template}.  The key point of the
technique lies in the remarkable similarity in the basic features
among galaxy spectra, although the relative intensity of absorption
lines can vary quite significantly, in particular when different
morphological types are considered.  In practice, to cover the range
of spectral properties a number of different templates is used for
each object and the one producing the highest cross-correlation peak
is then taken to be the best model, at the resulting redshift, of the
galaxy spectrum being measured. 

For measuring the spectra of the REFLEX survey, we benefited of the
accurate set of templates constructed by Ettori, Guzzo and Tarenghi
(1995, EGT hereafter), to which we refer for all details on their
properties and construction.  This set of templates has a number of
useful properties.  One advantage is that it includes separate stellar
and galaxy spectra together with composite spectra.  Another
important feature is the accurate knowledge of the template zero
points, calibrated in EGT using a set of high-resolution ``primary''
stellar templates.

This template library includes 17 spectra: two high-resolution HD stars with 
accurately known radial velocity, 3 high S/N galaxies observed with
EFOSC in a previous project (Collins et al. 1995), and combined
stellar and galaxy spectra built by EGT.

\subsubsection{Redshift Errors}
\label{errors}
The major advantage of the cross-correlation technique (Tonry \& Davis
1979) is to make use of the complete redshift information contained in
the whole spectrum, not just in the few major identifiable lines.
This pushes the measurement errors well below that expected from the
nominal spectral resolution used.  Depending on the SNR of the
spectrum, errors as small as 1/10 of the nominal accuracy on one
single-line measurement are achieved.  The specific IRAF
implementation \rvsao\/ computes a confidence level $R$ of the chosen
CCF peak as the ratio of the peak height to the {\it rms} background
of the CCF.  We empirically verified that estimates with $R<4$ have to
be treated with caution, while larger values normally indicated a
rather secure value.  We also used the stability of the redshift value
provided by the different templates as an extra figure of merit.  Each
galaxy spectrum was cross-correlated against the 17 templates
described above.  The overall results for each template were directly
inspected and spectra with 5 or more templates in agreement (within
the redshift errors) and $R>4$ were passed as secure.  Spectra that
did not satisfy these criteria strictly, had in several cases between
2 and 4 templates in agreement.  Visual inspection of these cases
often supported the suggested redshift.  The typical features of
cluster early-type galaxy spectra, as in particular the 4000 \AA~
break, make the visual check of the suggested redshift fairly
straightforward.  Spectra were discarded if (1) there was no agreement
between the templates and (2) the visual inspection did not indicate a
plausible redshift.  Once a spectrum had been accepted as secure
(visually or with $\geq5$ templates), the template redshift with the
highest $R$ parameter level was assigned to the galaxy.  If several
templates had the same confidence, then the one with the lowest
returned internal error was used.  For high signal-to-noise spectra,
it was common to find all the templates agreeing to within a scatter
of $\Delta v\simeq 50\,{\rm km\,s}^{-1}$.  The distribution of the
errors for the final redshifts is plotted in Fig.~\ref{err_hist}.
According to these histograms, the median formal error on our galaxy
redshifts is $\sim 60 \,{\rm km\,s}^{-1}$, with 70\% of them being
better than $100 \,{\rm km\,s}^{-1}$.  A small fraction of the
galaxies had emission lines in their spectra. These are indicated in
the redshift catalog, together with the corresponding emission-line
radial velocity.  This is normally of lower accuracy than the global,
cross-correlation based redshift which uses the information from the
whole absorption spectrum, and has been used to compute the cluster
redshift only when
no absorption redshift was available.  From our previous experience
with the same instrumental set-up (Collins et al. 1995), we also know
that the external error on the radial velocities (measured from
repeated observations of the same galaxies) is between a factor of one
and two larger than the {\tt XCSAO} internally estimated value (depending on
the SNR). This implies a median value for the total measurement errors
of $\sim 100\,{\rm km\,s}^{-1}$. 

\subsection{Galaxy Astrometry}

Precise astrometric coordinates were assigned {\it a posteriori} to
each observed spectrum, since they were available only approximately
from the header of the spectroscopic frames. For single-slit
observations this was in general straightforward, as (especially at
the two smaller telescopes) these involved fairly bright galaxies.
For MOS observations, on the other hand, it required a significant
amount of work, as unfortunately no electronic information on the
astrometric position of target galaxies on the MOS slits is saved
along with the observations at the telescope.  For this reason, we
calibrated astrometrically all the direct CCD images available for
each spectroscopic target field, using the USNO2 galaxy catalog and
the Starlink's Graphical Astronomy and Image Analysis Tool (GAIA, {\tt
  http://star-www.dur.ac.uk/pdraper/gaia/gaia.html}).  This was made
possible by using the service white-light CCD images used to prepare
the EFOSC slit masks, that were appropriately saved at the time of
observations.  For some fields, images in $B$ and $R$ bands were also
available.  
Inevitably, the final match of the 1D spectra and redshift to their
specific galaxy position on the sky was then performed by hand,
using the astrometrically calibrated images and the {\tt HEDIT} IRAF
task to write RA and DEC in the spectrum FITS header.

\section{CATALOG OF GALAXY REDSHIFTS AND OPTICAL 
DATA BASE}

\subsection{Galaxy Redshift Catalog}

During our spectroscopic observing campaign, we collected new
redshifts for 192 clusters which are included in the current REFLEX
catalog. Additionally, a number of systems with X-ray fluxes fainter
than the current REFLEX limit were also measured, together with
candidates that were subsequently discarded as non-cluster sources.
The full list of measured galaxy redshifts for clusters in the REFLEX
sample is provided in electronic form only (see {\tt
  www.brera.inaf.it/REFLEX}).  Here we provide only an excerpt, which
is displayed in Table~\ref{tab:gal_z}.  The columns give,
respectively: (1) REFLEX name, as defined in Paper II; 
(2,3) Coordinates J2000 of each target galaxy; (4) Simple spectral
classification, to distinguish among stars, galaxies and clear AGN-like
spectra. This classification is not meant to be
exhaustive. Additionally, when clear from the available imaging, the
spectroscopic measurement of the cD galaxy is
explicitly noted; (5) Assignment as a cluster member (+) or interloper
(-); (6) heliocentric redshift $cz$ in km s$^{-1}$, as measured from
absorption lines through the cross-correlation procedure; (7)
corresponding error; (8) redshift confidence parameter $R$, giving the
ratio between the height of the cross-correlation peak and the overall
rms noise of the same function; (9) emission-line redshift, when
available; (10,11) Observation date and telescope; (12) When needed,
notes on the spectrum quality or on possible problems in the
reduction, astrometry or redshift quality.
 
%



%


\subsection{Finding Charts and Optical/X-ray Atlas}

\begin{figure*}                                                                 
\resizebox{\hsize}{!}{\includegraphics{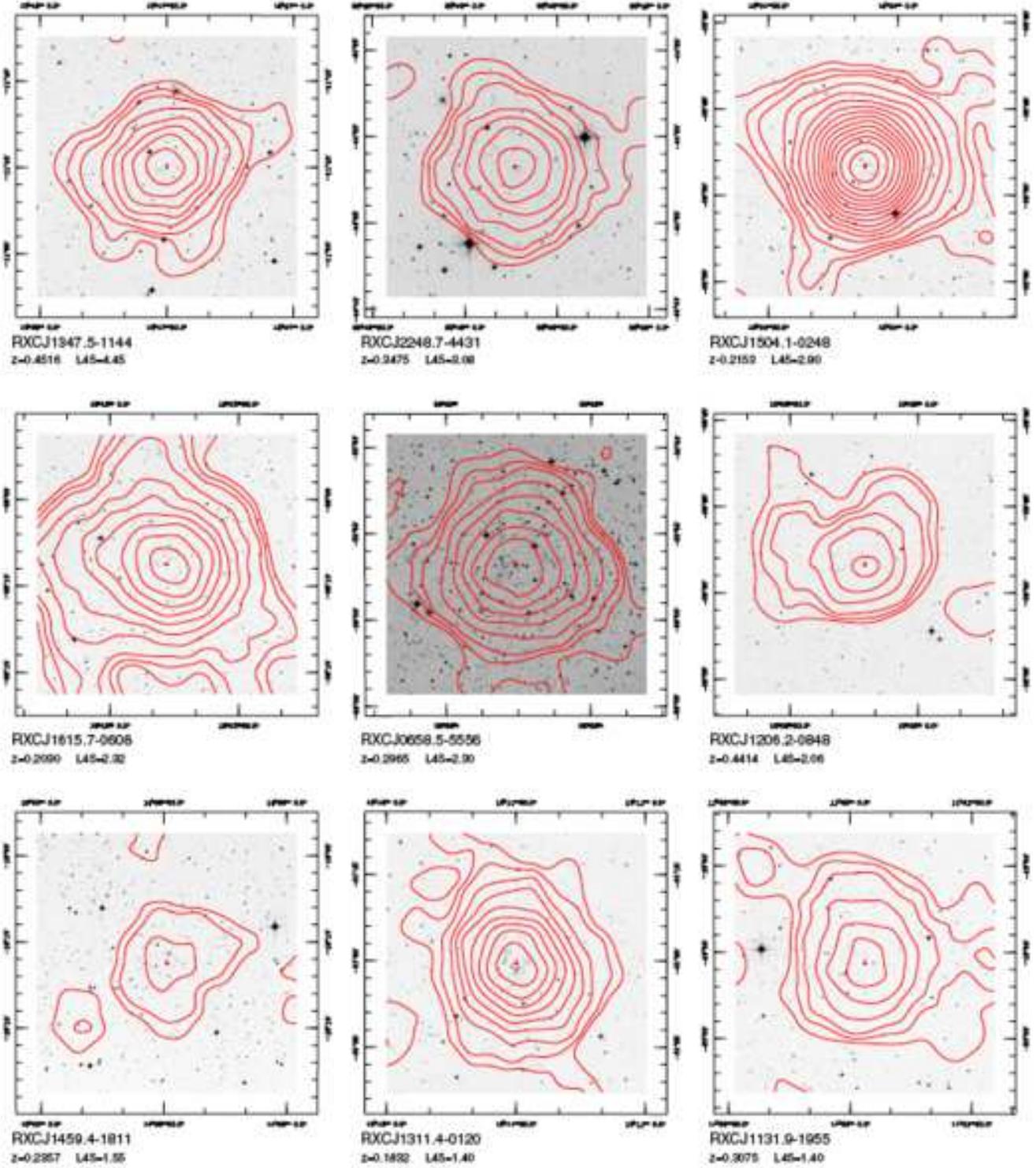}} 
\caption{Overlays of the X-ray emission in the [0.5-2.0]
  KeV band plotted onto the DSS2-RED images of clusters in the REFLEX surveys. 
  The contours correspond to
  steps of 1$\sigma$ in the {\it significance} of the X-ray emission,
  defined as the {\it rms} fluctuation within a Gaussian window
    of 1-arcmin-dispersion of the ratio $S/\sqrt{B+S}$, where $S$ is
  the source signal and $B$ the mean value of the background.  We show
 here the 9 most
  luminous systems in the survey. For each cluster, we also
  report its redshift and X-ray luminosity in units of $10^{45}$ erg
  s$^{-1}$ cm$^{-2}$ (as given in Paper II).  The full set of overlays
  and finding charts is available at higher resolution from the survey
  web page.   These clusters
  include objects already known from previously existing catalogs,
  as A2163 (RXCJ1615.7-0608), A1689 (RXCJ1311.4-0120) and A1300
 (RXCJ1131.9-1955) from the Abell catalog (Abell 1958) or 
S1063 (RXJ2248.7-4431) and S0780 (RXCJ1459.4-1811) from its
Supplementary list.  Notable is also RXCJ0658.5-5556, which
corresponds to the famous ``Bullet Cluster'' (Clowe et al.~2006)
originally discovered by the Einstein observatory as 1ES 0657-558.
} 
\label{overlays}
\end{figure*}  

\begin{figure*}
\resizebox{\hsize}{!}{\includegraphics{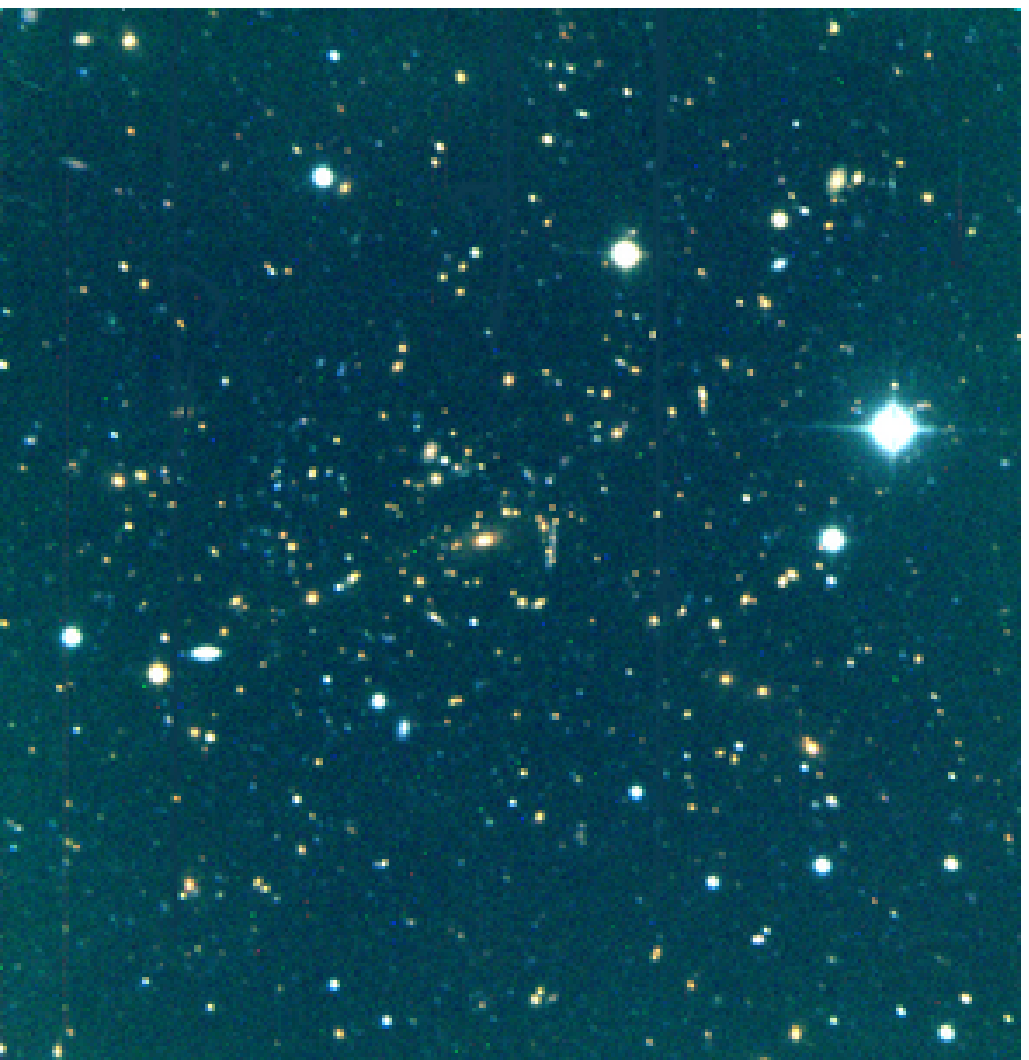}}
\caption{Composite RGB image of RXCJ1206.2-0848, one of the most
  spectacular new clusters discovered by the REFLEX survey (the sixth
  most luminous shown in Fig.~\ref{overlays}).  This image has been
  built combining three short (10 min) direct exposures in the Johnson $B$, $V$
  and $R$ bands taken with EFOSC2 at the ESO 3.6~m telescope and is
  5.5 arcmin on a side.  Note the
  dominating cD with very extended yellowish halo, and the prominent blue
  gravitational lensing candidate arc just westward of it. Several
  other possible arclets are also visible.
}
\label{fig:rgb}
\end{figure*}

A complete optical/X-ray atlas of images for the REFLEX clusters,
including finding charts for the spectroscopically measured galaxies
is too big to be included in this paper. We have therefore set up a
visual atlas of DSS finding charts and X-ray overlays, which is
accessible through the survey web page ({\tt
  www.brera.inaf.it/REFLEX}). The scientific content of the X-ray
overlays and their construction are discussed in a separate paper
(B\"ohringer et al., in preparation). The web page will also be
  used to present future upgrades of the REFLEX catalog, or new
  information on single clusters.

As a visual example of the most spectacular
objects which are part of the catalog, we show here a printed
version of the overlays for the nine most 
luminous REFLEX clusters  (Figure~\ref{overlays}). Some of these
  are famous clusters already known before REFLEX, as detailed in the
  caption. Some others are new objects discovered by REFLEX. These
  include, for example RXCJ1347.4-1144 at $z=0.4516$, the most luminous X-ray
  cluster known to date (Schindler et al. 1995). Another spectacular
  example of these newly discovered systems is RXCJ1206.2-0848 at
  $z=0.4414$, for which we show in Fig.~\ref{fig:rgb} an RGB composite
  of three CCD images in the $B$, $V$ and $R$ bands.  

We hope the easy-to-browse cluster imaging atlas will be useful for
planning specific studies of cluster sub-classes, as e.g. cD clusters
or the so-called ``fossil'' groups (of which REFLEX includes a
remarkable sub-set).

\section{Luminosity and Spatial Distribution}

The new redshifts for galaxies in REFLEX clusters presented here have
been used, together with a large bulk of existing data from the
literature, to assign a systemic redshift to each cluster, as
described in Paper II and to compute a velocity dispersion for a
sub-set of 170 objects, as reported in Ortiz-Gil et al. (2004) .  In
Paper II, we already presented a first discussion of the sample
resulting from the redshift survey, mostly concentrating on the
unambiguous identification of the redshift system related to the X-ray
source.  We briefly summarize some of these aspects here, presenting
some further details on the properties of the REFLEX cluster sample
and its spatial distribution.


%
\begin{figure}                                                                 
\resizebox{\hsize}{!}{\includegraphics{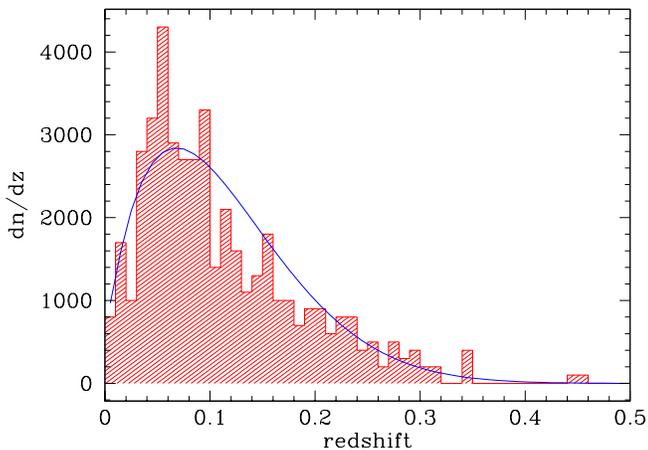}}
\caption{Redshift distribution of the REFLEX clusters (histogram),
compared to that expected from integration of the REFLEX X-ray
Luminosity Function from B\"ohringer et al. (2002).}
\label{nz}
\end{figure}  
\begin{figure}
\resizebox{\hsize}{!}{\includegraphics{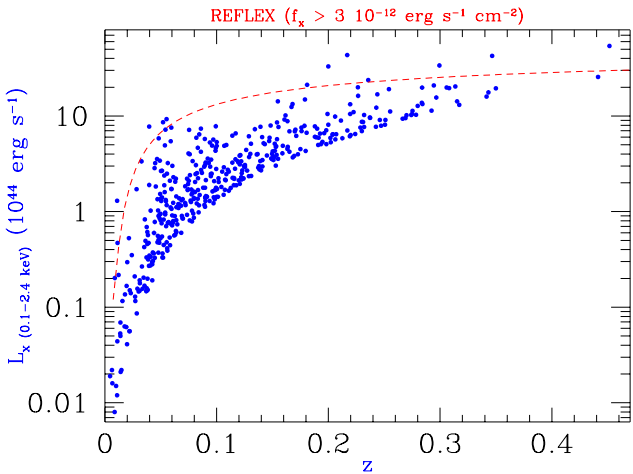}}
\caption{X-ray luminosity versus redshift for the final REFLEX sample of
galaxy clusters.  The lower cut-off in $L_X$ corresponds to the
survey flux limit of $3 \times 10^{-12}$ erg s$^{-1}$ cm$^{-2}$.  The
upper dashed line gives instead the expected volume effect imposed by
the luminosity function.  It is computed as the luminosity value above
which, at every redshift, less than 1 cluster is expected within the
enclosed cosmological volume.  This curve shows clearly how only with
a volume as big as that of REFLEX one can adequately explore the
bright end of the luminosity function.}
\label{Lx-z}
\end{figure}  

Figure~\ref{nz} shows the redshift distribution of the 447
clusters included in the REFLEX catalog. As a consistency check,
this is compared to the curve one obtains by integrating as a function
of redshift the no-evolution X-ray luminosity function (XLF hereafter)
measured from the sample itself (B\"ohringer et al. 2002).
Figure~\ref{Lx-z}, instead, plots the X-ray luminosity $L_X$ of the
clusters as a function of redshifts.  The plot shows how the
  REFLEX sample is able to include some of the most X-ray luminous
  clusters in the Universe, thanks to its large volume.   It is
evident how all very luminous systems with
$L_X>10^{45}$ erg s$^{-1}$ are found above $z>0.15$. This is the
consequence of these clusters being rare fluctuations lying on the
exponential tail of the luminosity function: at any redshift, there is
a maximum luminosity $L_{MAX}$ , above which the expected number of
clusters (given by the integral of the luminosity function $\phi(L)$ above
$L_{MAX}$ times the volume explored), drops below unity. Following
Sandage, Tammann \& Yahil (1979), the value of $L_{MAX}$ as a function of
redshift is implicitly provided by the expression
\begin{equation}
N(<z, >L_{MAX}) =  1  \, ,
\end{equation}
i.e.
\begin{equation}
\int_0^{V(z)} dV \int_{L_{MAX}}^\infty dL\; \phi(L) = 1  \,.
\end{equation}

The corresponding solution $L_{MAX}(z)$, given the REFLEX XLF
corresponds to the dashed
curve in Figure~\ref{Lx-z}.  The curve describes fairly well the
upper envelope of the $L_X-z$ plot, with fluctuations around it
produced by large-scale structures (where the mean density, and thus
the normalization $\phi^*$ of the XLF fluctuates around the mean value
used in the computation).

\begin{figure*}
\resizebox{\hsize}{!}{\includegraphics{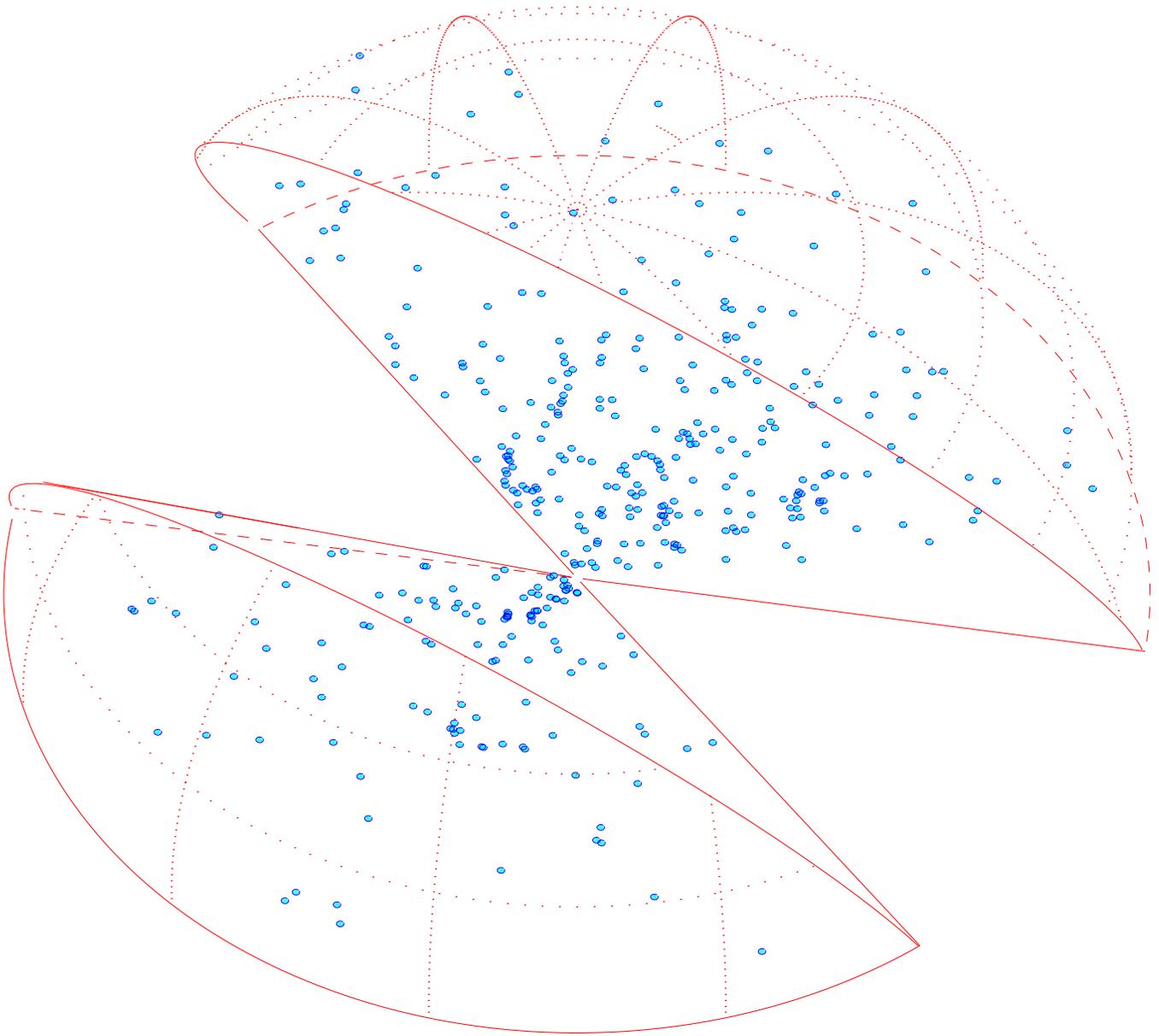}}
\caption{The large-scale spatial distribution of REFLEX clusters
within 600 $\hmpc$.  The South Galactic Pole is here placed on top to
ease display.  The missing wedge is the region occupied by the
Galactic plane ($\pm 20^\circ$).
}
\label{fig:coneplot}
\end{figure*}

Finally, Figure~\ref{fig:coneplot} provides an overview of the 3D
distribution of REFLEX clusters, within $z<0.2$.  One can easily
notice the level of structure still existing on such very large
scales, with a number of evident aggregations of clusters with sizes
$\sim 100 \hmpc$.  

%

\section{SUMMARY}

The REFLEX survey consists of 447 galaxy clusters constituting the
largest statistically complete (to better than 90 \%) X-ray
flux-limited cluster survey to date. The 
spectroscopic follow-up of REFLEX was carried out as part of an ESO
Key Programme using a combination of single slit and multi-object
spectroscopy providing new redshifts for 1406 galaxies in these
systems. Clusters were observed with either the Faint Object
Spectrograph and Camera on the 2m and 3.6m ESO telescopes or the
Boller and Chivens spectrograph on the 1.5m. These combinations
provide a spectral wavelength coverage of between $3600 - 8000\AA$ and
a two-pixel resolution of $\simeq 14 \AA$. Redshifts are measured mainly
by cross-correlation with a range of template spectra.
Internal fitting errors and external comparisons indicate that galaxy redshifts
are typically accurate to 100 km s$^{-1}$, with errors as small as 50
km s$^{-1}$ for the highest SNR spectra.  We have produced
optical/X-ray overlays for all clusters, together with finding charts
indicating the spectroscopically observed galaxies. These are
available at  {\tt
  http://www.brera.inaf.it/REFLEX}, together with the complete table
of all galaxy redshifts.

\begin{acknowledgements}                                                        
  We would like to thank G. Chincarini, R.~Cruddace, H.~MacGillivray,
  T.~Reiprich, K. Romer, W.~Seitter, P.~Vettolani, W.~Voges
  and G.~Zamorani for 
  their contribution and support to various aspects of the REFLEX
  project.  The help of U.~Briel, R. D\"ummler, H.~Ebeling,
  G.~Guerrero, D.~Lazzati, E.~Molinari, S.~Molendi and D.~Wake during some
  observing sessions is also gratefully acknowledged.  We thank A. Mist\`o
  for the development and maintenance of the Brera database system,
  which has been vital to support parts of this work.
  The redshift survey of REFLEX clusters was
  performed in the framework of the ESO ``Key Programmes''.  We would
  like to acknowledge the generous allocation of telescope time by ESO
  OPC.  We thank the ESO-La Silla staff for invaluable support, and
  in particular F. Patat and M. Kurster for special assistance during
  observations at the 3.6~m telescope.  We thank the referee, Harald
  Ebeling, for his comments that helped to improve the presentation of
  the paper.   
LG acknowledges financial support from MIUR (through grant PRIN 98
``Galaxy Formation and Evolution''), the Italian Space Agency (ASI) and
the hospitality of the Excellence Cluster ``Universe'' in Garching
(supported by DfG grant EXC 153), where this paper was completed.
AOG has been supported by the Spanish MEC project AYA2003-08739-C02-01
(including FEDER) and by the Generalitat Valenciana ACyT project
GRUPOS03/170. 

This research has made use of the
NASA/IPAC Extragalactic Database (NED)  
which is operated by the Jet Propulsion Laboratory, California Institute 
of Technology, under contract with the National Aeronautics and 
Space Administration. It has also made use of NASA's Astrophysics 
Data System.

This paper and the whole REFLEX project would not have been possible
without the dedication of Peter Schuecker, who passed away in
November 2006.  He has been the driving force behind both the
observations and the key scientific results obtained by the REFLEX
survey.  We all miss his theoretical knowledge, pure approach to science and
unique humanity. 

\end{acknowledgements}


\begin{thebibliography}{}

\bibitem{} 
Abell, G.O., 1958, ApJS, 3, 211

\bibitem{} 
Abell, G.O., Corwin, H.G. \& Olowin, R.P., 1989, ApJS, 70, 1

\bibitem{}
Allen, S., MNRAS, 2001, 328, 37 

\bibitem{}
Avila, G., D’Odorico, S., Tarenghi, M. \& Guzzo, L., 1989, The Messenger, 55, 62. 

\bibitem{}
B\"ohringer, H., Voges, W., Huchra, J.P., McLean, B.,
Giacconi, R., Rosati, P., Burg, R., Mader, J.,
Schuecker, P., Simi{\c c}, D., Komossa, S.,
Reiprich, T.H., Retzlaff, J., Tr\"umper, J., 2000, ApJS, 129, 435

\bibitem{}
B\"ohringer, H., Schuecker, P., Guzzo, L., Collins, C.A., 
Voges, W., Schindler, S., Neumann, D.M., Cruddace, R.G., De Grandi, S.,
Chincarini, G., Edge, A.C., MacGillivray, H.T., Shaver, P., 2001,
A\&A, 369, 826 (Paper I)

\bibitem{}
B\"ohringer, H., Schuecker, P., Komossa, S., Retzlaff, J., Reiprich, T.H.,
\& Voges, 2001b, in {\it Mapping the Hidden Universe},
R.C. Kraan-Korteweg, P.A. Henning 
\& H. Andernach (eds.), p. 93, astro-ph/0011461

\bibitem{}  
B\"ohringer, H., Collins, C.A., Schuecker, P., Guzzo, L.,
Voges, W., Neumann, D.M., Schindler, Cruddace, R.G., De Grandi, S.,
Chincarini, G., Edge, A.C., Reiprich, T.H., \& Shaver, P., 2002, ApJ,
566, 93  

\bibitem{} 
B\"ohringer, H., Schuecker, P., Guzzo, L., Collins, C.A., 
Voges, W., Cruddace, R.G., Ortiz-Gil, A., De Grandi, S.,
Chincarini, G., Edge, A.C., MacGillivray, H.T., Neumann, D.M.,
Schindler, S., Shaver, P., 2004,  
A\&A, 425, 367  (Paper II)

\bibitem{} 
B\"ohringer, H., Schuecker, P., Pratt, G.W., Arnaud, M., Ponman, T.J.,
et al., 2007, A\&A, 469, 363  

\bibitem{} 
Borgani, S., \& Guzzo, L., 2001, Nature, 409, 39

\bibitem{} 
Borgani, S., Murante, G., Springel, V. et al., 2004, MNRAS, 348, 1078 

\bibitem{} 
Borgani, S., Rosati, P., Tozzi, P., et al., 2001, ApJ, 561, 13

\bibitem[Burns et al.(1996)]{1996ApJ...467L..49B} 
Burns, J.~O., Ledlow, 
M.~J., Loken, C., Klypin, A., Voges, W., Bryan, G.~L., Norman, M.~L., \& 
White, R.~A.\ 1996, ApJ, 467, L49

\bibitem{}
Clowe, D., Bradac, M., Gonzalez, A.H., Markevitch, M., Randall, S.W.,
Jones, C., Zaritsky, D., 2006, ApJ, 648, L109

\bibitem{}
Collins, C.A., Guzzo, L., Nichol, R.C., \& Lumsden, S.L.,
1995, MNRAS, 274, 1071

\bibitem{} 
Collins, C.A., Guzzo, L., B\"ohringer, H., Schuecker, P.,
Chincarini, G., Cruddace, R., De Grandi, S., Neumann, D., Schindler,
S., \& Voges, W., 2000, MNRAS, 319, 939

\bibitem{}
Crawford, C.S., Allen, S.W.,  Ebeling, H., Edge, A. C., Fabian, A.C.
1999, MNRAS, 306, 857

\bibitem{}
Cruddace, R., Voges, W., B\"ohringer, H., Collins, C.A., Romer, K.A., 
MacGillivray, H.T., Yentis, D., Schuecker, P., Ebeling, H., 
De Grandi, S., 2002, ApJS, 140, 239

\bibitem{}
De Grandi, S., Molendi, S., B\"ohringer, H., \& Voges, W., 1997,
ApJ, 486, 738  

\bibitem{}
De Grandi, S., B\"ohringer, H., Guzzo, L., et al., 1999,
ApJ, 514, 148

\bibitem{}
Ebeling, H., Voges, W., B\"ohringer, H., Edge, A.C.,
Huchra, J.P., Briel, U.G., 1996, MNRAS, 281, 799

\bibitem{}
Ebeling, H., Edge, A.C., B\"ohringer, H., et al., 1998, MNRAS, 301, 881

\bibitem{}
Ebeling, H., Edge, A.C., Allen, S.W., Crawford, C.S., Fabian, A.C.,
\& Huchra, J.P., 2000, MNRAS, 318, 333

%
%

\bibitem{}
Ettori, S., Guzzo, L., Tarenghi, M., 1995, MNRAS, 276, 689 (EGT)

\bibitem{}
Ettori, S., Tozzi, P., Borgani, S., Rosati, P., 2004, A\&A, 417, 13   

\bibitem{}
Evrard, G., Metzler, C.A., \& Navarro, J.F., 1996, ApJ, 469, 494  

\bibitem{}
Gioia, I.M., Henry, J.P., Mullis, C.R., et al. 2003, ApJS, 149, 29

\bibitem{}
Finoguenov, A., Reiprich, T.H., \& B\"ohringer, H., 2001, A\&A, 368, 749

\bibitem{}
Guzzo, L., B\"ohringer, H., Schuecker, P., et al., 1999, The Messenger, No. 95, 27

\bibitem{}
Helsdon, S.F, \& Ponman, T.J., 2000, MNRAS, 319, 933

\bibitem{}
Henry, J.P., Gioia, I.M., Mullis, C.R., et al., 2001, ApJ, 553, L109

\bibitem[Henry(2003)]{2003mecg.conf....5H} 
Henry, J.~P.\ 2003, in ``Matter and 
Energy in Clusters of Galaxies'', ASP Conf. Proc., Vol.~301,  
S. Bowyer and C.-Y. Hwang eds, San Francisco, p.~5

\bibitem{}
Heydon-Dumbleton, N., Collins, C.A., \& MacGillivray, H.T., 1989, MNRAS, 238, 379 

\bibitem{}
Kaiser, N., 1986, MNRAS, 222, 323   

\bibitem{}
Kerscher, M., Mecke, K., Schuecker, P., B\"ohringer, H., Guzzo, L., 
Collins, C. A., Schindler, S., De Grandi, S., Cruddace, R., 2001, 
A\&A, 377, 1

\bibitem{}
Kurtz, M.J., and Mink, D.J., 1998, PASP, 110, 934

\bibitem[Ledlow et al.(1999)]{1999ApJ...516L..53L} 
Ledlow, M.~J., Loken, 
C., Burns, J.~O., Owen, F.~N., \& Voges, W.\ 1999, ApJ, 516, L53 
 
\bibitem{}
MacGillivray, H.T. \& Stobie, R.S., 1984, Vistas Astr., 27, 433

\bibitem{}
Maddox, S. J.; Efstathiou, G.; Sutherland, W. J.; Loveday, J. , 1990,
MNRAS 243, 692 

\bibitem{}
Olowin, R., De Souza, R.E., Chincarini, G., 1988, A\&AS, 73, 125

\bibitem{} 
Ortiz-Gil, A., Guzzo, L., Schuecker, P., B\"ohringer, H., Collins,
C.A., 2004, MNRAS, 348, 325  

\bibitem{}
Pierpaoli, E. et al., 2003, MNRAS, 342, 163, 

\bibitem{}
Pierre, M., et al., 1994, A\&A, 290, 725    

\bibitem{}
Reiprich T.H. \& B\"ohringer, H., 2002, ApJ, 567, 716

\bibitem{}
Romer, A.K., Collins, C.A., B\"ohringer, H., Cruddace, R.G.,
Ebeling, H., MacGillivray, H.T., \& Voges, W., 1994, Nature, 372, 75

\bibitem{}
S\'anchez, A.G., Lambas, D.G., Böhringer, H., Schuecker, P., 2005,
MNRAS, 362, 1225

\bibitem{}
Sandage, A., Tammann, G.A., \& Yahil, A., 1979, ApJ, 232, 352

\bibitem{}
Schindler, S., Guzzo, L., Ebeling, H., Boehringer, H., Chincarini, G.,
Collins, C. A., de Grandi, S., Neumann, D. M., Briel, U. G., Shaver,
P., Vettolani, G., 1995, A\&A, 299, 9 

\bibitem{}
Schindler, S., Hattori, M., Neumann, D. M., Boehringer, H.,1997, A\&A,
317, 646   

\bibitem{}
Schuecker, P., B\"ohringer, H., Guzzo, L., Collins, C.A., Neumann, D.M.,
Schindler, S., Voges, W., Chincarini, G., Cruddace, R.G., De Grandi, S.,
Edge, A.C., M\"uller, V., Reiprich, T.H., Retzlaff, J., \& Shaver, P.,
2001, A\&A, 368, 86

\bibitem{}
Schuecker, P., Guzzo, L., Collins, C.A., B\"ohringer, H., 2002, 
MNRAS, 335, 807 

\bibitem{}
Schuecker, P., B\"ohringer, H., Collins, C.A., Guzzo, L., 2003a,
A\&A, 398, 867

\bibitem{}
Schuecker, P., Caldwell, R.R., B\"ohringer, H., Collins, C.A., Guzzo,
L., Weinberg, N.N., 2003b, A\&A, 402, 53,

\bibitem{}
Stanford, S.A., et al., 2006, ApJ, 646, 13

\bibitem{}
Stanek, R., et al., 2006, ApJ, 648, 956

\bibitem{}
Struble, M.F. \& Rood, H.J. 1999, ApJS, 125, 35

\bibitem{}
Tonry, J.L., and Davis, M., 1979, AJ, 84, 1511

\bibitem{}
Voges, W., Aschenbach, B., Boller, T., Br\"auninger, H.,
Briel, U., Burkert, W., Dennerl, K., Englhauser, K., Gruber, R.,
Haberl, F., Hasinger, G., K\"urster, M., Pfeffermann, E.,
Pietsch, W., Predehl, P., Rosso, C., Schmitt, J.H.M.M., Tr\"umper, J.,
\& Zimmermann, H.U., 1999, A\&A, 349, 389


\end{thebibliography}
\end{document}